\documentclass[12pt]{article}
\PassOptionsToPackage{svgnames,x11names,table,dvipsnames}{xcolor}

\usepackage{amssymb,amsmath,amsfonts,geometry,graphicx,caption,setspace,footmisc,pdflscape,%
  array,hyperref,booktabs,threeparttable,titlesec}
\usepackage{xcolor}
\usepackage{natbib}

\hypersetup{
    colorlinks=true,
    linkcolor=blue,
    citecolor=blue,
    filecolor=magenta,
    urlcolor=blue,
    pdftitle={Ownership and Non-Neutral Technological Change: Evidence from China's State-Owned Enterprise Privatization},
    pdfauthor={Ziyao Wang},
    pdfpagemode=FullScreen,
    }

\begin{document}

\begin{titlepage}
\title{Ownership and Non-Neutral Technological Change: Evidence from China's State-Owned Enterprise Privatization\thanks{I am grateful to my dissertation committee, James Dana, Jordi Jaumandreu, Santiago Caicedo, and Jianfei Cao, for their invaluable guidance. I am especially indebted to Jordi Jaumandreu for many discussions. I also thank Yan Bai, John Kwoka, Jun Ma, Matthias Mertens, Bettina Peters, Wing Suen, Heng Yin, and Shunan Zhao for their insightful suggestions. I thank Jordi Jaumandreu and Heng Yin for their help with the dataset. I appreciate helpful comments from participants at the Northeastern Economics Workshop, BU Micro Workshop, BU--BC Green Line Meeting, ZEW, CES-2025, and WEAI-2025 conferences. A previous version circulated under the title ``Non-neutral Technological Change in Chinese Manufacturing". Any remaining errors are my sole responsibility.
}}
\author{
  Ziyao Wang\thanks{
    Ph.D. Candidate in Economics, Northeastern University.
    Email: \href{mailto:wang.ziyao1@northeastern.edu}{wang.ziyao1@northeastern.edu}
  }
}

\date{This version: \today\\ First version: March 2025}
\maketitle

\vspace{-0.1cm}

\begin{abstract}
\singlespacing
\noindent I estimate firm-level capital-, labor-, and material-augmenting productivity for Chinese manufacturing, 1998--2008, and the effect of state-owned enterprise (SOE) privatization on each. Results show technological change is strongly biased toward capital and labor. Private firms operate above SOEs on capital and labor, but below on materials. Privatization raises labor-augmenting productivity right after the transition, a front-loaded gain that fades on the full panel but persists among 1998 incumbents. It leaves capital-augmenting productivity no higher and modestly lowers materials. The SOE sector's catch-up runs through restructuring of the plants the state kept, with exit of overstaffed plants adding to the labor gain.
\\
\vspace{0in}

\end{abstract}
\setcounter{page}{0}
\thispagestyle{empty}
\end{titlepage}
\pagebreak \newpage

\onehalfspacing

\interfootnotelinepenalty=10000

\section{Introduction} \label{sec:introduction}

Beginning in the late 1990s, China restructured, privatized, or closed most of its state-owned enterprises (SOEs). Privatization is expected to raise productivity, but the empirical evidence is contested: \citet{hsieh2015grasp} conclude that the privatization of state firms contributed little to aggregate growth, yet \citet{chen2021privatization} find that privatization raises total factor productivity (TFP) among the 1998 cohort of incumbents. Estimating the effect of the ownership change is challenging because privatization was selective and the firms the state privatized and those it kept were both improving rapidly during the reform decade. The state sector entered the reform with a factor-specific legacy of underused capital, redundant labor, and administratively allocated materials, which a single TFP index cannot distinguish. The question is whether technological change at privatized firms outpaced that at comparable firms the state kept, and which factor it augmented.

This paper estimates firm-level capital-, labor-, and material-augmenting productivity for Chinese manufacturing from 1998 to 2008, and the effect of SOE privatization on each. I find SOEs stand below non-state firms in capital- and labor-augmenting productivity, though above them in materials, and they close much of that gap over the reform decade. The act of privatization itself raises labor-augmenting productivity in the early years after the transition, as overstaffed state firms shed surplus workers, but yields no lasting capital-augmenting gain. Most of the state sector's catch-up comes from the improvement of the SOEs the state kept and, for labor, the exit of the most overstaffed plants.

To recover the three separately, I propose a new method to estimate the CES production function with three factor-augmenting productivities, which is robust to product market power. The estimator starts from the firm's static cost-minimization problem. Combining the first-order conditions with the observed input cost shares lets me rewrite the CES technology as three single-factor output equations, each involving only one input, its cost share, and its factor-augmenting productivity. Treating productivity as an AR(1) process augmented with time effects and ownership controls, I estimate the system jointly by dynamic-panel GMM \citep{arellano1991some,blundell2000gmm}, with lagged inputs and cost shares as instruments, recovering the elasticity of substitution, the returns to scale, and the persistence of each factor-augmenting productivity. A Generalized Least Squares (GLS)-based filter then separates each firm's productivity path from measurement error.

To estimate the ownership-level and privatization effects, I model each factor-augmenting productivity as a controlled Markov process whose law of motion depends on the firm's ownership type and privatization status, as in the endogenous productivity literature of \citet{doraszelski2013r}, \citet{deloecker2013detecting}, \citet{braguinsky2015acquisitions}, and \citet{chen2021privatization}. Against a baseline of the SOEs the state kept, the process carries ownership-type terms for never-state private, collective, and foreign firms and for exiting SOEs, and, for the durable privatizers, separate pre- and post-privatization terms, with the post term split by destination (SOE to private or foreign, and SOE to collective) and by timing (the early and late years after the transition).

To address the privatization selection behind ``grasp the large, let go of the small,'' under which the state privatized or closed firms non-randomly, I benchmark privatizers against the SOEs the state kept and exploit the reform's timing. Because privatization and liquidation took eight to fourteen months from decision to implementation, tied to the state's annual property-rights review, a firm's realized ownership in a given year is set by the previous year's information and is predetermined with respect to its current-period productivity shock, following \citet{chen2021privatization}. Predeterminedness makes the realized transition indicators valid instruments, and a destination-matched pre-transition control absorbs selection on the pre-privatization productivity level, so no structural selection model is required.

The results show that technological change is strongly biased toward capital and labor: capital-augmenting productivity grows fastest, at about $16.9\%$ a year, closely followed by labor at $14.2\%$, while material-augmenting productivity is nearly flat at $0.4\%$, with an elasticity of substitution of $0.384$ and returns to scale near constant. The raw cost shares point the same way. Over the decade the capital cost share of Chinese manufacturing fell by nearly two-thirds, and the fall was concentrated among SOEs and privatizing firms rather than non-state firms.

Ownership status shapes the level of this productivity. Non-state firms stand well above SOEs in capital- and labor-augmenting productivity, in line with the TFP literature \citep{brandt2012creative,hsieh2015grasp,chen2021privatization}, but I find the ordering reverses for materials, where SOEs are the most productive, a new pattern that likely reflects state firms' preferential access to intermediate inputs through administrative allocation. Among the SOEs themselves, the plants the state let go were the least productive of all, far below the ones it kept.

The act of privatization, by contrast, works through labor and, on the full panel, only at the transition. When privatizers are compared before and after their transition against the kept SOEs, so that the selection of which firms were privatized is differenced out, privatization raises labor-augmenting productivity significantly in the early years after the transition, a front-loaded gain that fades thereafter, while it yields no lasting capital-augmenting gain and modestly reduces material-augmenting productivity. An event study of the recovered productivity corroborates the structural timing for labor, tracing the same rise-then-plateau path on flat pre-transition leads.\footnote{See the event study in \ref{Appendix:eventstudy}.} Capital shows the same transitory bump then decline, though a pre-trend in its leads leaves the capital path only suggestive. Averaged over all post-transition years the labor gain is small, matching the small aggregate effect of \citet{hsieh2015grasp}. It is larger and more lasting within the incumbent 1998 cohort that \citet{chen2021privatization} study: these incumbents carry the heaviest legacy of administrative overstaffing, so privatization lets them shed more surplus labor and the gain persists well beyond the transition window, while the full panel dilutes it with later, leaner SOE entrants. In both samples the privatization effect runs through labor.

The state sector nonetheless closes much of its capital- and labor-augmenting gap with private firms over the decade. It closes the capital gap mostly within the plants the state kept, which improve faster than any other group on a capital stock they barely expand while shedding excess labor, with a further composition margin from state firms entering the sample and no lift from exit. It closes the labor gap through those same kept plants and entrants, aided by the exit of the most overstaffed ones. The reform's productivity gains thus run through selection and the restructuring of the plants the state kept rather than through the transformation of privatized firms, whose factor-augmenting productivity, beyond the transition-window labor gain, stays close to the kept-SOE trajectory and short of the always-private firms it never reaches. Privatization is instead associated with higher markups, in line with \citet{chen2021privatization}'s emphasis on pricing and demand management.\footnote{See the markup analysis in \ref{Appendix:markup}.}

The results also differ sharply across industries. Capital-augmenting growth ranges from about $10\%$ a year in textiles to above $30\%$ in metals, and the capital bias is strongest in several of the heavy industries that entered the reform with large stocks of old equipment. Labor-augmenting growth varies less, and material-augmenting growth stays near zero everywhere. The privatization effect is small in most industries, with the labor gains concentrated in the transition window and in the most overstaffed sectors: the only positive early capital effect appears in metals, where privatizers entered furthest below the kept SOEs and closed part of the gap, and the clean early labor gains appear in machinery and transport and in chemicals, the sectors where administrative overstaffing was heaviest. The exit margin, by contrast, differs widely across sectors. The SOEs the state let go sit furthest below the kept ones in exactly the industries where a simple level comparison of privatized firms against the remaining state pool, a pool dominated by those exiters, would have shown the biggest apparent gains.

This article contributes to three literatures. First, it adds to the work measuring non-neutral technological change at the firm level. Building on the directed technical change framework of \citet{acemoglu1998new,acemoglu2002directed}, recent studies recover factor-specific productivity from firm data: \citet{doraszelski2018measuring} estimate labor-augmenting and Hicks-neutral efficiency for Spanish manufacturing, \citet{raval2019micro} estimates labor-augmenting productivity for U.S.\ plants, and \citet{demirer2020production} recovers it under a weaker functional-form assumption than CES to study output elasticities and markups.\footnote{Ongoing work by \citet{zhao2024estimation} adopts a translog form with labor-augmenting and neutral productivity.} \citet{zhang2019non} extends the factor efficiency measure to capital, labor, and materials for the Chinese steel industry, starting from profit maximization under monopolistic competition and estimating it by control-function methods.\footnote{\citet{gao2023firm} apply a similar framework across Chinese industry.} I contribute to this literature in two ways. The first is identification of all three factor-augmenting productivities. I recover them from cost-minimization conditions and estimate them by dynamic-panel GMM, so the results are robust to firms' market power,\footnote{\citet{jaumandreu2024robust} discusses the two main approaches to estimating firm-level production functions, the dynamic-panel methods \citep{arellano1991some,blundell2000gmm} and the control-function methods \citep{olley1996dynamics,levinsohn2003estimating,ackerberg2015identification}, and considers the dynamic-panel approach robust to market power, holding under perfect, monopolistic, or oligopolistic competition.} and no product market structure constrains the markup.\footnote{The trade-off to recover the capital-augmenting productivity is that \citet{zhang2019non}'s demand elasticity implies an industry-level constant markup, which is valid in the relatively homogeneous steel industry, while my approach assumes no product-market structure but treats capital as a flexible input. I provide a robustness check for the latter, allowing a capital adjustment-cost wedge, in Section~\ref{sec:capwedge}.} The second is scope. I let ownership and privatization enter the productivity process, so I can measure the effect of ownership change on each of the three factor-augmenting productivities. The new findings show that ownership types differ in their levels of factor efficiency and that a change of ownership, such as privatization, has heterogeneous effects across the three.

Second, the article contributes to the literature on endogenous productivity, in which the law of motion of productivity depends on firm states or choices: \citet{aw2011rd} and \citet{doraszelski2013r} model R\&D as a determinant of productivity, \citet{deloecker2013detecting} lets export status shift the Markov process, and \citet{braguinsky2015acquisitions} study productivity around ownership changes in Japanese cotton spinning. \citet{chen2021privatization} apply this framework to Chinese SOE privatization and model a Hicks-neutral productivity. I extend the controlled-Markov approach to a factor-augmenting CES technology, letting ownership change shift capital-, labor-, and material-augmenting productivity separately, so I can ask whether privatization is factor-biased instead of summarizing it in one index. The two designs also differ in what pins down the post-privatization level. In \citet{chen2021privatization}, ownership enters through current-type terms, so a privatized firm is assumed to converge to the productivity level of the private type. Here the post-transition terms are parameters of their own, estimated separately from the never-state terms, and the privatization effect is identified from the privatizers' own pre- and post-transition years, measured against the SOEs.

Third, the article adds new findings to the literature on how China's SOE reform and privatization affect productivity. Whether ownership change raises productivity has been contested. On one side, \citet{brandt2012creative} find state-owned firms less productive than other firms, and \citet{chen2021privatization} find private firms more productive than SOEs, with gains that accrue over several years. On the other, \citet{hsieh2015grasp} find that the exit and privatization of SOEs had a negligible effect on aggregate growth.\footnote{\citet{hsieh2015grasp} also report firm-level labor and capital productivity, but these are average products (value added per worker and per unit of capital) rather than factor-augmenting efficiencies. Their structural productivity is a single Hicks-neutral TFP index in a Cobb--Douglas technology. } My factor-level estimates reconcile these views. In levels, and consistent with the TFP literature \citep{brandt2012creative,chen2021privatization}, I find non-state firms stand well above SOEs in capital- and labor-augmenting productivity, but a new finding is that non-state firms are below SOEs in materials.\footnote{\citet{brandt2012creative} and \citet{hsieh2015grasp}'s total factor productivity is measured in value added, which nets materials out, so this material-augmenting reversal lies outside what such single-index comparisons capture.} How productivity moves with privatization then depends on the sample. The small average post-transition labor effect on the full panel echoes \citet{hsieh2015grasp}, while the 1998-cohort labor gain matches the population \citet{chen2021privatization} study, and my estimation adds that the gain from privatization runs mostly through labor.

The remainder of the article is organized as follows. Section~\ref{sec:data} describes the data and the ownership classification. Section~\ref{sec:model} sets up the CES model and the controlled-Markov productivity process, and Section~\ref{sec:Estimation} the estimation. Section~\ref{sec:result} reports the pooled estimates and the privatization effects, Section~\ref{sec:heterogeneity} industry heterogeneity, and Section~\ref{sec:discussion} the decomposition of the state sector's catch-up. Section~\ref{sec:capwedge} assesses robustness to capital adjustment frictions, and Section~\ref{sec:conclusion} concludes.

\section{Chinese Manufacturing Context} \label{sec:data}

\subsection{Data}

I use the dataset cleaned by \citet{jaumandreu2018cost}, based on the Annual Census of Industrial Production conducted by the National Bureau of Statistics (NBS) of China\footnote{I use \citet{jaumandreu2018cost}'s whole cleaned panel, not the exporter subsample they construct, in order to retain full coverage of non-exporters and of state-owned and collective firms.}. The survey comprises all industrial non-state firms with more than 5 million RMB in annual sales plus all industrial state-owned firms (SOEs). My analysis focuses on data from 1998 to 2008 because the SOE reform and privatization mainly happened in this decade and the NBS dataset post-2008 faces several issues, including missing information on value-added and intermediate inputs, as well as concerns about data consistency for my study \citep{brandt2023has}.

The resulting analysis panel consists of $372{,}813$ firms and $1{,}659{,}612$ firm-year observations spanning $1998$--$2008$. I classify firms into ten sectors based on their industry codes. I construct variables for production estimation, including revenue, output price, capital stock, user cost of capital, wage bill, employment, material price, and material cost. I also add an ownership indicator based on the financial capital structure. Summary statistics for these variables are reported in~\ref{Appendix:data}, which also provides detailed descriptions of the variables and outlines the correspondence between industries and the two-digit code classification.

\subsection{Privatization and Ownership Classification} \label{sec:priv_data}

A central feature of my sample is the wave of state-owned enterprise restructuring and privatization during 1998--2008. The reform operated on three margins: the internal restructuring of the state firms the government retained, the exit of the weakest through liquidation or closure, and the privatization or
collectivization of others. Privatization is thus only one component of the reform, a distinction I maintain throughout, so that the privatization effects estimated below concern that channel and not the reform's overall productivity gains. Over this period, the number of majority state-owned firms in the panel fell from approximately $17{,}000$ to $3{,}300$, a decline of more than $80\%$. This selective survival, ``grasp the large, let go of the small'' policy, motivates my endogenous productivity treatment of ownership transitions in Section~\ref{sec:model}.

Following \citet{brandt2012creative} and \citet{hsieh2015grasp}, I classify firms into four ownership types in each year based on the identity of the majority shareholder, using a $50\%$ equity threshold to determine control: state-owned enterprises (SOEs), private firms, collective enterprises, and foreign-invested firms.\footnote{Two approaches coexist in the literature for defining ownership types: a registration-based classification (e.g., \citealp{chen2021privatization}) and a shareholder-based classification using a majority-control threshold (e.g., \citealp{brandt2012creative,hsieh2015grasp}). I adopt the latter, which uses the shareholding threshold to capture the degree of state control in Chinese firms.} A firm is classified as an SOE in year $t$ if its state paid-in capital share is at least $50\%$. The other three categories follow analogously. This year-by-year classification allows firms to switch ownership types over time. It therefore picks up the ownership transitions of the reform decade. Firms with no majority shareholder (``mixed'' ownership) account for less than $1\%$ of firm-years and are not separately analyzed. Panel~A of Table~\ref{tab:ownership_counts} reports the resulting ownership composition over the sample period. The number of SOEs falls from $17{,}271$ in $1998$ to $3{,}351$ in $2008$, while the number of private firms rises from $29{,}146$ to $189{,}337$.

My sample contains substantial ownership transitions during this decade, tracked at the firm-id level by following each firm's state-capital share across years.\footnote{Privatization in China usually changes a firm's registration type and hence its NBS identifier. The cleaned panel of \citet{jaumandreu2018cost} therefore links records across identifier changes, following \citet{brandt2012creative} (matching on firm name, address, and other identifiers), as do \citet{chen2021privatization} and \citet{hsieh2015grasp}.} Table~\ref{tab:priv_stats} summarizes the resulting classification. Of the $35{,}663$ firms ever classified as SOE, $16{,}843$ are majority state-owned throughout their panel spell. Of these, only $1{,}608$ survive as SOEs to 2008 (the ``kept'' SOEs that form the omitted baseline of my productivity process), while $15{,}235$ exit the panel before 2008, the low-productivity plants the reform ``let go.'' A further $11{,}998$ firms make a durable SOE-to-non-SOE transition: a first downcrossing of the $50\%$ threshold with no subsequent reversion. I restrict the privatization sample to the $8{,}695$ clean privatizers: firms that are SOE at their first observation and leave state ownership exactly once. This excludes two groups that the naive first-downcrossing rule would mislabel as SOE privatizations: $3{,}303$ ``re-nationalizers'' (private firms briefly classified as SOE and then re-privatized, whose return to private status is not a privatization of a state firm), which I retain in the estimation sample but assign to a non-interpreted nuisance category. The second group is $3{,}613$ ``flicker'' firms that cross the threshold more than once, which I drop. A residual $3{,}209$ firms are nationalized (private or collective at first observation, later crossing into majority state ownership without a durable privatization). They are ever-SOE but outside the privatization sample, and with this group the partition exhausts the $35{,}663$ ever-SOE firms. Each clean privatizer is assigned a destination, private/foreign or collective, by the ownership type it settles into after the transition, defined as the type it holds for most of its post-transition years. The $52$ privatizers that settle into mixed ownership (no majority shareholder, Table~\ref{tab:priv_stats}) are grouped with the private/foreign destination. Measured against the kept SOEs, these durable privatizers form the identifying variation that underlies my productivity process in Section~\ref{sec:model}.

\begin{table}[htbp]
\centering
\begin{threeparttable}
\caption{Ownership Composition and the Privatization Sample.}
\label{tab:ownership_counts}%
\label{tab:priv_stats}%
{\small
\setlength{\tabcolsep}{4.5pt}
\begin{tabular}{l r r r r r r}
\toprule
 & SOE & Private & Collective & Foreign & Mixed & Total \\
\midrule
\multicolumn{7}{l}{\textit{Panel A: Number of firms by ownership type and year}} \\
\addlinespace
1998 & 17{,}271 & 29{,}146  & 25{,}128 & 4{,}166  & 238 & 75{,}949  \\
1999 & 19{,}092 & 38{,}441  & 27{,}462 & 5{,}041  & 300 & 90{,}336  \\
2000 & 16{,}193 & 48{,}168  & 23{,}971 & 5{,}618  & 247 & 94{,}197  \\
2001 & 13{,}112 & 62{,}589  & 20{,}666 & 6{,}529  & 216 & 103{,}112 \\
2002 & 11{,}134 & 76{,}587  & 17{,}842 & 7{,}774  & 192 & 113{,}529 \\
2003 & 9{,}056  & 95{,}365  & 14{,}045 & 9{,}208  & 150 & 127{,}824 \\
2004 & 7{,}272  & 142{,}758 & 13{,}234 & 14{,}424 & 125 & 177{,}813 \\
2005 & 6{,}090  & 163{,}200 & 10{,}715 & 15{,}844 & 98  & 195{,}947 \\
2006 & 5{,}155  & 187{,}315 & 9{,}734  & 17{,}319 & 110 & 219{,}633 \\
2007 & 4{,}168  & 213{,}713 & 9{,}051  & 18{,}817 & 121 & 245{,}870 \\
2008 & 3{,}351  & 189{,}337 & 6{,}400  & 16{,}229 & 85  & 215{,}402 \\
\addlinespace
\multicolumn{6}{l}{\textit{Distinct firms in panel, 1998--2008}} & 372{,}813 \\
\midrule
\multicolumn{7}{l}{\textit{Panel B: Privatization sample (firm-level counts)}} \\
\addlinespace
Ever-SOE firms                & & & & & & 35{,}663 \\
\quad Kept SOEs (to 2008)\tnote{$\dagger$}     & & & & & & 1{,}608  \\
\quad Exited SOEs (pre-2008)  & & & & & & 15{,}235 \\
\quad Clean privatizers       & & 7{,}325 & 1{,}078 & 240 & 52 & 8{,}695 \\
\quad Excluded (re-nat.\ + flicker) & & & & & & 6{,}916 \\
\quad Became-SOE (nationalized)\tnote{$\ddagger$} & & & & & & 3{,}209 \\
\bottomrule
\end{tabular}
}
\begin{tablenotes}[flushleft]
\footnotesize
\item \textit{Notes:} Panel~A classifies a firm under an ownership type if that type's paid-in capital share is at least $50\%$; yearly counts are cross-sectional and exceed the number of distinct firms because most firms appear in multiple years. Panel~B tracks ownership at the firm-id level.
\item[$\dagger$] Kept SOEs are state-owned in every year and survive to 2008; clean privatizers appear under the Panel~A column of their durable post-transition destination.
\item[$\ddagger$] Became-SOE ($3{,}209$): private or collective at first observation, later crossing into majority state ownership without a durable privatization; ever-SOE but outside the privatization sample.
\end{tablenotes}
\end{threeparttable}
\end{table}

\subsection{Stylized Facts on Ownership and Factor-Biased Technological Change} \label{sec:stylized}

\paragraph{Non-neutral technological change in Chinese manufacturing.}
The raw data already point to strongly factor-biased rather than Hicks-neutral technological change. Panel~A of Table~\ref{tab:stat_whole_data} reports firm-level averages of inputs and outputs in Chinese manufacturing in 1998 and 2008. The number of firms rose sharply, and the average firm grew larger: deflated revenue roughly tripled and material use more than doubled. Against this expansion, employment per firm fell steadily, by more than a third, while real capital stayed roughly flat. That output rose so much faster than either labor or capital is a first sign that firms were getting more out of each unit of these two inputs.

Panel~B of the same table reports the three factor cost shares. The capital cost share fell sharply, from $10.3\%$ to $3.8\%$. The labor cost share fell only modestly, from $11.4\%$ to $10.3\%$. The material cost share rose, from $78.3\%$ to $86.0\%$.

\begin{table}[htbp]
\centering
\begin{threeparttable}
\caption{Chinese Manufacturing: Firm-Level Averages and Factor Cost Shares, 1998 and 2008.}
\label{tab:stat_whole_data}
\begin{tabular}{lcc}
\toprule
 & \textbf{1998} & \textbf{2008} \\
\midrule
\multicolumn{3}{l}{\textit{Panel A: Firm-level averages}} \\
No.\ of firms              & 75{,}949 & 215{,}402 \\
Revenue\tnote{a}           & 43.7     & 120.5     \\
Capital\tnote{b}           & 28.0     & 27.7      \\
Labor (workers)            & 333      & 211       \\
Materials\tnote{c}         & 32.5     & 79.7      \\
\midrule
\multicolumn{3}{l}{\textit{Panel B: Factor cost shares in total cost}} \\
Capital share              & 0.103    & 0.038     \\
Labor share                & 0.114    & 0.103     \\
Material share             & 0.783    & 0.860     \\
\bottomrule
\end{tabular}
\begin{tablenotes}
\footnotesize
\item[a] Deflated by industry output price indices; millions of RMB.
\item[b] Deflated by an investment price index; millions of RMB.
\item[c] Deflated by industry materials price indices; millions of RMB.
\item \textit{Notes:} Panel B reports unweighted firm-mean cost shares (input cost over total cost).
\end{tablenotes}
\end{threeparttable}
\end{table}

Cost shares and factor productivity are tightly linked. When a firm becomes more productive at using an input, it needs less of that input relative to the others. If the inputs are hard to substitute for one another, its expenditure share on that input falls. A steep decline in a factor's cost share therefore signals a particularly large improvement in how productively that factor is used.

A natural candidate for these factor-biased changes is the contemporaneous SOE reform: the restructuring, privatization, and exit of underperforming state plants may have let the firms deploy their inputs more productively than before.

\begin{figure}[htbp]
\centering
\includegraphics[width=\textwidth]{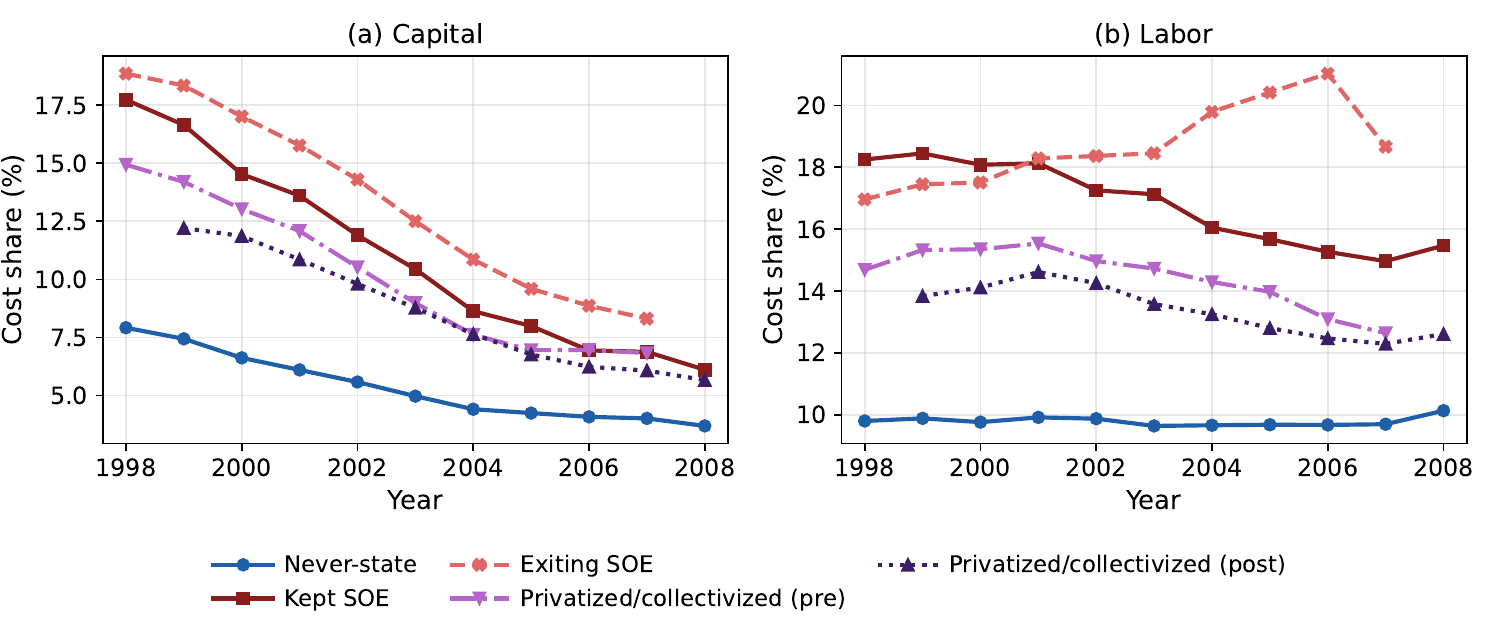}
\caption{Factor cost shares by ownership group, 1998--2008 (unweighted firm mean). Panel~(a) reports the capital cost share $S^K=P^K K/TC$ and panel~(b) the labor cost share $S^L=P^L L/TC$. Groups follow the firm-level ownership histories defined in Section~\ref{sec:priv_data}. The privatized/collectivized group is the clean durable SOE$\to$non-SOE transitions (Table~\ref{tab:ownership_counts}), pooling privatization (to private/foreign) and collectivization (to collective control) in the sense of \citet{chen2021privatization}. It appears post-transition only from 1999 (the first year any firm transitions), and exiting SOEs are observed only through their last year.}
\label{fig:cost_shares}
\end{figure}

\paragraph{Ownership and non-neutral technological change.} \label{sec:share_by_own}
The decline in the capital cost share $S^K=P^K K/TC$ is overwhelmingly a state-sector phenomenon (Figure~\ref{fig:cost_shares}, panel~(a)). Kept SOEs enter the period with a capital share more than twice that of never-state firms and converge down toward it by 2008, durable privatizers sit in between and fall as they leave state ownership, and never-state firms barely move. The labor cost share shows the same state versus non-state divide (panel~(b)): it declines among kept SOEs and privatizers and stays flat among never-state firms, while it rises among the exiting SOEs on their way out (by a small sample size at the end). The factor-share signal of biased technical change to capital and labor is thus concentrated in the firms the reform touched.

The pattern suggests that ownership and privatization shaped the direction of factor-biased change as well as its pace, but this reading is tentative. These are raw group averages, so they mix the reform's effect with how firms sort into and out of state ownership. To separate what actually happens to a firm when it is privatized from these pre-existing group differences, I turn to the structural model of Section~\ref{sec:model}, which builds ownership and a post-privatization indicator into how each factor's productivity evolves and estimates these effects directly.

\section{Empirical Model} \label{sec:model}

I begin from a Constant Elasticity of Substitution (CES) production function with firm-level capital-, labor-, and material-augmenting productivity, and show that combining the cost-minimization first-order conditions with the observed cost shares rewrites it as three single-factor output equations.

\subsection{Production Function} Consider a Constant Elasticity of Substitution production function of firm $j$ at time $t$ with non-neutral technology:

\begin{equation}
    Q_{jt} = \left\{ \left[\exp\left(\omega^K_{jt}\right) K_{jt} \right]^{-\frac{1-\sigma}{\sigma}}
    + \left[\exp\left(\omega^L_{jt}\right) L_{jt} \right]^{-\frac{1-\sigma}{\sigma}}
    + \left[\exp\left(\omega^M_{jt}\right) M_{jt} \right]^{-\frac{1-\sigma}{\sigma}} \right\}^{-\frac{\sigma}{1-\sigma}\nu}.
\label{eq:CES}
\end{equation}
The terms $\omega^{X}_{jt}$ represent the firm-specific factor-augmenting productivities, which are the parameters of interest, where $X$ denotes one of the three input factors: capital ($K$), labor ($L$), or materials ($M$). These factor-augmenting productivities are observed by firms at the time of production but remain unobserved by researchers. The parameter $\nu$ denotes the elasticity of scale and $\sigma$ the elasticity of substitution.

The observed output, measured as firm revenue divided by the product price index, is
\mbox{$Y_{jt} = Q_{jt} \cdot \exp(\varepsilon_{jt})$}, where $\varepsilon_{jt}$ represents a measurement error that is uncorrelated across firms and over time, with a mean of zero.

\subsection{Deriving Factor-Share Equations from Cost Minimization}
I start from the firm's cost-minimization problem rather than profit maximization. The resulting input first-order conditions depend only on observable input prices and technology, not on unobserved markups or demand elasticities, so I can estimate the production parameters without any assumption on the firm's behavior in the product market.\footnote{Because every firm (monopolist, oligopolist, or price taker) must still choose the cheapest mix of inputs for whatever output level it ultimately sells, the first-order conditions from conditional cost minimization equate each input price to marginal cost times its marginal product independently of how the firm later sets price or markups. That separation means the cost-share equations I exploit remain valid under perfect competition, oligopoly, or monopoly, so long as input markets are competitive.}

I assume that capital, labor, and materials are static inputs chosen each period to minimize production costs.\footnote{Although capital is a durable asset, the classical user-cost argument of \citet{hall1967tax} and \citet{martin1984misuse} implies that a firm facing depreciation, financing, and opportunity costs behaves as if it rented capital services each period at the prevailing user cost. At the annual frequency of my data, plants adjust those services by purchasing, scrapping, leasing, or varying utilization, so the capital first-order condition holds under the user-cost formulation.} Treating capital as static follows several factor-augmenting studies \citep{raval2019micro,oberfield2021micro,gao2023firm} and reflects the central trade-off of my approach. Requiring static cost minimization for capital, as well as for labor and materials, is what frees the estimator from any assumption on product-market competition and so allows for various market power in the product market. Over this decade of rapid Chinese restructuring, when firms were actively buying, scrapping, and reallocating capital, it could be a reasonable assumption. I test it directly in Section~\ref{sec:capwedge}, where allowing a capital adjustment cost wedge between the user cost and the shadow price of capital leaves the structural estimates unchanged and does not overturn the privatization conclusions.\footnote{A uniform wedge is absorbed by the capital-year effects, while an ownership-specific state-capital subsidy can only lower the capital privatization effect below its baseline estimate, which is therefore an upper bound.}

The firm is further assumed to be a price taker in the input market \citep{doraszelski2018measuring, raval2019micro, zhang2019non, demirer2020production}, facing $P^K_{jt}$, $P^L_{jt}$, and $P^M_{jt}$ as the prices of capital, labor, and materials, respectively. Applying the FOCs from cost minimization, together with the envelope theorem, yields:

\begin{equation}
	 MC_{jt} \frac{\partial Q_{jt}}{\partial X_{jt}} = P^X_{jt},
\end{equation}
where \( MC_{jt} \) denotes the marginal cost of production.

To eliminate the unobserved marginal cost term, I consider the ratio of the first-order conditions for labor and materials. Multiplying both sides of the labor and material FOCs by \( \frac{L_{jt}}{M_{jt}} \) yields:
\begin{equation}
\left[\frac{\exp(\omega^L_{jt}) L_{jt}}{\exp(\omega^M_{jt}) M_{jt}}\right]^{-\frac{1-\sigma}{\sigma}} = \frac{P^L_{jt} L_{jt}}{P^M_{jt} M_{jt}}	.
\label{eq4}
\end{equation}

I then define the total cost of firm $j$ at time $t$ as $TC_{jt}= P^K_{jt} K_{jt} + P^L_{jt} L_{jt} + P^M_{jt} M_{jt}$, and the factor cost share as $S^X_{jt} = P^X_{jt}X_{jt}/TC_{jt}$. This definition of input share deviates from the approach used in \citet{zhang2019non} and \citet{gao2023firm}, where input share is defined as the ratio of cost to revenue. I adopt input cost divided by total cost as input share to avoid potential endogeneity issues arising from revenue-based shares.\footnote{The summation of input shares in revenue incorporates both the elasticity of scale and demand elasticity, which introduce additional sources of endogeneity. However, input shares in total cost, as used in this article, which sum to one by construction, are directly observable, and do not depend on demand and product market structure assumptions.}

Using the definition of cost shares, Equation (\ref{eq4}) can be rearranged as:
\begin{equation}
	\frac{\exp(\omega^L_{jt}) L_{jt}}{\exp(\omega^M_{jt}) M_{jt}} = \bigg(\frac{S^L_{jt}}{S^M_{jt}} \bigg)^{-\frac{\sigma}{1-\sigma}}.
    \label{eq_cost_share}
\end{equation}

Analogous expressions can be derived for capital-labor and capital-materials pairs. Substituting these share-based expressions into the CES production function allows me to rewrite output as a function of a single input factor, its corresponding cost share, and its input-specific productivity. This yields a system of three equations:

\begin{equation}
\left\{
\begin{aligned}
& Q_{jt} =   \left[\exp(\omega^K_{jt}) K_{jt} \right] ^\nu \left(\frac{1}{S^K_{jt}}\right)^{-\nu \frac{\sigma}{1-\sigma}} \\
& Q_{jt} =   \left[\exp(\omega^L_{jt}) L_{jt} \right] ^\nu \left(\frac{1}{S^L_{jt}}\right)^{-\nu \frac{\sigma}{1-\sigma}} \\
& Q_{jt} =   \left[\exp(\omega^M_{jt}) M_{jt} \right] ^\nu \left(\frac{1}{S^M_{jt}}\right)^{-\nu \frac{\sigma}{1-\sigma}}
\end{aligned}
\right. .
\end{equation}

Given that the observed output has an error term, $Y_{jt} = Q_{jt} \cdot \text{exp}(\varepsilon_{jt})$, I take the natural logarithm of both sides. Following standard convention, I use lowercase letters to denote logarithmic values:

\begin{equation}
\left\{
\begin{aligned}
& y_{jt} = \nu  k_{jt} +\nu \frac{\sigma}{1-\sigma}  s^K_{jt} + \nu  \omega^K_{jt}  + \varepsilon_{jt} \\
& y_{jt} = \nu  l_{jt} +\nu \frac{\sigma}{1-\sigma}  s^L_{jt} + \nu  \omega^L_{jt}  + \varepsilon_{jt} \\
& y_{jt} = \nu  m_{jt} +\nu \frac{\sigma}{1-\sigma}  s^M_{jt} + \nu  \omega^M_{jt}  + \varepsilon_{jt}
\end{aligned}
\right. .
\label{eq:beforegmm}
\end{equation}

\subsection{Productivity Process} \label{sec:prodprocess}
I model each factor-augmenting productivity $\omega^X_{jt}$ as a controlled Markov process whose law of motion depends on the firm's ownership state and privatization status, following \citet{aw2011rd}, \citet{doraszelski2013r}, \citet{deloecker2013detecting}, \citet{braguinsky2015acquisitions}, and \citet{chen2021privatization}. It is an AR(1) driven by a deterministic shifter $z^X_{jt}$,
\begin{equation}
\omega^X_{jt} = \rho^X \omega^X_{j,t-1} + z^X_{jt} + \xi^X_{jt},
\qquad X \in \{K,L,M\},
\label{eq:ar1}
\end{equation}
where $\rho^X$ is the persistence of factor $X$, common to all firms. The shifter $z^X_{jt}$ is the state-dependent component of the conditional expectation of period-$t$ factor-augmenting productivity, through which ownership status shifts the law of motion. I date the ownership indicators at $t$, following \citet{chen2021privatization}.\footnote{\citet{doraszelski2013r} and \citet{deloecker2013detecting} assume that R\&D or export status in $t-1$ are predetermined, so these are uncorrelated with the innovation shock. Similarly, \citet{chen2021privatization} and I assume that the ownership type and the privatization decision are predetermined at the beginning of the year $t$.} It collects the year fixed effect and the ownership controls. Let $\mathrm{NPri}_{jt}$, $\mathrm{NCol}_{jt}$, $\mathrm{NFor}_{jt}$ denote indicators for never-SOE firms currently classified as private, collective, and foreign-controlled. The indicator $\mathrm{ExitSOE}_{jt}$ marks firms majority state-owned throughout their panel spell that exit before 2008, and $\mathrm{PreSOE}^{SP}_{jt}$ and $\mathrm{PreSOE}^{SC}_{jt}$ mark the still-SOE years of firms that are later durably privatized, split by their eventual destination. The indicators $\mathrm{POST}^{d,w}_{jt}$ mark a durable transition out of majority state ownership by $t$, split by destination $d$, to private or foreign control ($SP$) or to collective control ($SC$),\footnote{The two destinations essentially exhaust the durable privatizers: in my sample about $88\%$ settle into private or foreign control and $12\%$ into collective control. The $52$ firms that settle into mixed ownership are grouped with the private/foreign destination (Section~\ref{sec:priv_data} gives the assignment rule). The collective destination is the one \citet{chen2021privatization} flag as a possible ``collectivization as euphemism'' for privatization. Unlike their design, which excludes foreign firms, I group the private and foreign destinations together (not separate SOE-foreign because it is a small sample). Reclassifications among the non-SOE types (in particular collective-to-private) are not counted, since they involve no change in state control.} and by timing window $w$, early ($\mathrm{ea}$) for the first three post-transition years ($s\equiv t-\tau_j\in\{0,1,2\}$, where $\tau_j$ is the transition year) and late ($\mathrm{la}$) for $s\geq3$, following the acquisition-window split of \citet{braguinsky2015acquisitions}. Splitting the post term this way relaxes the restriction that a single coefficient imposes, that the productivity effect of privatization is constant over all post-transition years.\footnote{The transition year $s=0$ is placed in the early window. Because ownership at $t$ is predetermined by the prior year's decision, $\mathrm{POST}^{d,\mathrm{ea}}_{jt}$ is orthogonal to $\xi^X_{jt}$ like the other transition indicators.}

Collecting these,
\begin{equation}
\begin{aligned}
z^X_{jt} \equiv\ {} & \gamma^X_t + \alpha^X_P\,\mathrm{NPri}_{jt} + \alpha^X_C\,\mathrm{NCol}_{jt} + \alpha^X_F\,\mathrm{NFor}_{jt} + \alpha^X_E\,\mathrm{ExitSOE}_{jt} \\
& + \pi^X_{SP}\,\mathrm{PreSOE}^{SP}_{jt} + \pi^X_{SC}\,\mathrm{PreSOE}^{SC}_{jt} \\
& + \sum_{d\in\{SP,SC\}}\ \sum_{w\in\{\mathrm{ea},\mathrm{la}\}} \beta^{X,w}_{d}\,\mathrm{POST}^{d,w}_{jt} + \eta^X\,\mathrm{Other}_{jt}, \qquad X\in\{K,L,M\}.
\end{aligned}
\label{eq:z_def}
\end{equation}
The innovation $\xi^X_{jt}$ has mean zero and is mean-independent of lagged productivity and of the firm's ownership indicator $z^X_{jt}$. The transition indicators $\mathrm{POST}^{d,w}_{jt}$ are predetermined with respect to $\xi^X_{jt}$ by the eight-to-fourteen-month implementation lag of Chinese state-asset restructuring, as in \citet{chen2021privatization}, and the ownership-type indicators are likewise predetermined up to their never-SOE condition, because they are functions of the firm's realized ownership state at $t$. Because $\mathrm{ExitSOE}$ and $\mathrm{PreSOE}^{d}$ classify a firm by the eventual outcome of its restructuring rather than by a state known at $t$, they rest on group membership that is fixed over the firm's spell. $\mathrm{ExitSOE}$ is on in every year of an exiting SOE's spell, while $\mathrm{PreSOE}^{d}$ is on only in the still-SOE years of a firm whose eventual destination is $d$, switching off at the transition when the $\mathrm{POST}^{d,w}$ indicators take over.\footnote{\citet{braguinsky2015acquisitions} likewise place pre- and post-acquisition indicators in the law of motion of productivity.} For these labels, mean independence requires the innovations to average to zero within each restructuring group. The state's exit and privatization choices may depend on the productivity levels and ownership types of the firms it selects, differences the level terms absorb and estimate, but not on the current innovation $\xi^X_{jt}$. Conditional on the group, productivity follows the AR(1) of Equation~(\ref{eq:ar1}). The same requirement covers the never-SOE condition of the ownership-type indicators. The no differential trend of future privatizers relative to kept SOEs before the transition is tested by the leads of the event study in \ref{Appendix:eventstudy}.

Realized factor-augmenting productivity thus decomposes into a component expected under the firm's path, $\rho^X\omega^X_{j,t-1}+z^X_{jt}$, which ownership status and the privatization indicator are allowed to shift, and the innovation $\xi^X_{jt}$. Omitting these ownership terms would instead force the persistent SOE-to-private trend into the innovation $\xi^X_{jt}$ and bias the persistence and technology parameters, the controlled-Markov analogue of the bias \citet{doraszelski2013r} and \citet{deloecker2013detecting} document from imposing an exogenous productivity process.

The ownership indicators partition all firm-years against a single omitted category: kept-SOE firm-years, firms majority state-owned throughout their spell that survive to 2008.  This is the policy-relevant counterfactual (the SOEs the state chose to ``grasp'').  Pulling the exiting SOEs out of the baseline through $\mathrm{ExitSOE}$ is essential. Majority-state firms that exit the panel outnumber kept SOEs by roughly five to one in firm-years and are far less productive, so the pooled ``currently-SOE'' baseline used implicitly elsewhere is dominated by low-productivity exiters and mechanically inflates every level term measured against it.\footnote{The exiting SOEs remain in the estimation sample throughout. Pulling them out of the baseline changes which group the coefficients are measured against, not which firm-years are used, so the survival conditioning of the kept-SOE group is a normalization rather than a sample restriction.  Nor does it drive the estimated privatization effect. Re-estimating the system with the $\mathrm{ExitSOE}$ control dropped and the exiting SOEs folded into the baseline leaves every wedge $\beta^{X,w}_d-\pi^X_d$ within $0.004$ of its estimate, while the level terms shift as expected against the lower pooled baseline. \ref{Appendix:baseline} reports the full comparison and explains the invariance.} The ownership-type terms $\alpha^X_P, \alpha^X_C, \alpha^X_F$ shift the expected productivity of each never-SOE type relative to kept SOEs, and $\alpha^X_E$ records the term of the exiting SOEs, a structural counterpart to the negative selection.  Like the R\&D and export terms of \citet{doraszelski2013r} and \citet{deloecker2013detecting}, these coefficients are per-period terms of the law of motion. They shift the conditional expectation of period-$t$ productivity. A firm that remains in a given state settles at a productivity level $\frac{\text{coefficient}}{(1-\rho^X)}$ log points from the kept-SOE baseline, the distances the group trajectories of Section~\ref{sec:priv_effect} display.\footnote{For two firms sharing the year effects, a permanent difference $\alpha$ in the shifter compounds to a level gap $g$ solving $g=\rho^X g+\alpha$, that is, $\alpha/(1-\rho^X)$. The text reports these implied steady-state gaps where levels are the object of interest.}  The coefficients $\pi^X_{SP}, \pi^X_{SC}$ on $\mathrm{PreSOE}$ test for selection on trajectory. They measure how privatizers of each destination differ from kept SOEs in expected factor-$X$-augmenting productivity before they privatize.  The post-transition coefficients $\beta^{X,w}_{d}$ give the terms relative to kept SOEs by destination $d$ and window $w$. The privatization effect on factor-augmenting productivity is the baseline-invariant wedge $\beta^{X,w}_{d}-\pi^X_{d}$, the shift in the conditional expectation of period-$t$ factor-$X$-augmenting productivity attributable to leaving state ownership in window $w$, net of the destination-matched pre-privatization differential $\pi^X_d$ that controls for selection of SOEs into privatization.  A residual ``mixed/messy'' group (never-SOE firms with no majority shareholder, together with ever-SOE firms whose ownership histories are not clean, chiefly re-nationalizations) is absorbed by the nuisance indicator $\mathrm{Other}_{jt}$, whose coefficient $\eta^X$ is estimated but not interpreted, so that the kept-SOE baseline stays pure.

I then plug Equation~(\ref{eq:ar1}) into Equation~(\ref{eq:beforegmm}) and take pseudo-differences by using $y_{jt} - \rho^X y_{jt-1}$, which generates a dynamic representation. I denote the composite error term as $\widetilde{\xi}^X_{jt}$, where $\widetilde{\xi}^X_{jt} = \nu \xi^X_{jt} + (\varepsilon_{jt} - \rho^X \varepsilon_{jt-1})$. I obtain the system of equations:

\begin{equation}
\left\{
\begin{aligned}
& y_{jt} = \rho^K y_{jt-1} + \nu (k_{jt} - \rho^K k_{jt-1}) +\nu \frac{\sigma}{1-\sigma} (s^K_{jt} - \rho^K s^K_{jt-1}) + \nu z^K_{jt} + \widetilde{\xi}^K_{jt} \\
& y_{jt} = \rho^L y_{jt-1} + \nu (l_{jt} - \rho^L l_{jt-1}) +\nu \frac{\sigma}{1-\sigma} (s^L_{jt} - \rho^L s^L_{jt-1}) + \nu z^L_{jt} + \widetilde{\xi}^L_{jt} \\
& y_{jt} = \rho^M y_{jt-1} + \nu (m_{jt} - \rho^M m_{jt-1}) +\nu \frac{\sigma}{1-\sigma} (s^M_{jt} - \rho^M s^M_{jt-1}) + \nu z^M_{jt} + \widetilde{\xi}^M_{jt}
\end{aligned}
\right. .
\label{eq:gmm}
\end{equation}

\section{Estimation} \label{sec:Estimation}

I estimate the system in two steps. Equation~(\ref{eq:gmm}) is estimated by the Generalized Method of Moments (GMM), which delivers the elasticity of substitution ($\sigma$), the elasticity of scale ($\nu$), the persistence parameters ($\rho^K, \rho^L, \rho^M$), and the coefficients of the shifter $z^X_{jt}$ in Equation~(\ref{eq:z_def}). Those estimates turn Equation~(\ref{eq:beforegmm}) into three panel sequences that mix latent productivity with measurement error, as in Equation~(\ref{eq:residulas}). Moment matching identifies the variances of the two, and a Generalized Least Squares (GLS) filter then recovers each firm's productivity path.

\subsection{GMM} Equation~(\ref{eq:gmm}) comprises multiple equations, and there are common coefficients among different equations, which are $\nu$ and $\sigma$. This setting requires me to adopt the Generalized Method of Moments (GMM) estimator of \citet{hansen1982large}.

Following the current assumptions in the estimating firm productivity literature, I assume that firms choose inputs after observing current-period productivity \citep{arellano1991some, olley1996dynamics, blundell2000gmm, levinsohn2003estimating,ackerberg2015identification}, so that the contemporaneous input $x_{jt}$ and its cost share $s^X_{jt}$ are correlated with the productivity shock $\xi^X_{jt}$, while their lagged values are not. I therefore use lagged input usage and cost shares as instruments, as they are uncorrelated with the composite error in Equation~(\ref{eq:gmm}). The ownership components of the deterministic shifter $z^X_{jt}$ instrument their own coefficients under the timing and path arguments of Section~\ref{sec:prodprocess}.

The following moment conditions are used to estimate the structural parameters:
\begin{equation}
\left\{
\begin{aligned}
\mathbb{E}\left[\left(s^{L}_{jt-1}\right)'\widetilde{\xi}^L_{jt}\right]=0\\
\mathbb{E}\left[\left(s^{M}_{jt-1}\right)'\widetilde{\xi}^M_{jt}\right]=0\\
\mathbb{E}\left[\left(k_{jt-1}\right)'\widetilde{\xi}^K_{jt}\right]=0\\
\mathbb{E}\left[\left(l_{jt-1}\right)'\widetilde{\xi}^L_{jt}\right]=0\\
\mathbb{E}\left[\left(m_{jt-1}\right)'\widetilde{\xi}^M_{jt}\right]=0
\end{aligned}
\right. ,
\label{eq:instruments}
\end{equation}
where $s^{X}_{jt-1}$ represents the logarithm of the lagged cost share of factor $X$, while $x_{jt-1}$ denotes the logarithm of the lagged usage of factor $X$. Specifically, I have two instruments that are the lagged cost shares ($s^{L}_{jt-1}, s^{M}_{jt-1}$), and three instruments that are the lagged input quantities ($k_{jt-1}, l_{jt-1}, m_{jt-1}$). These five instruments provide just-identification of the five structural parameters $(\sigma, \nu, \rho^K, \rho^L, \rho^M)$.

The deterministic shifter $z^X_{jt}$ collects the year fixed effects $\gamma^X_t$, the ownership-type terms $\alpha^X_P, \alpha^X_C, \alpha^X_F, \alpha^X_E$, the two destination-specific pre-privatization selection terms $\pi^X_{SP}, \pi^X_{SC}$, the four destination-and-timing privatization coefficients $\beta^{X,w}_{d}$, and the nuisance coefficient $\eta^X$. These coefficients are identified by treating each component of the shifter as its own instrument, under the orthogonality arguments of Section~\ref{sec:prodprocess}. Stacking the three factor equations, which share the dependent variable $y_{jt}$ and the structural parameters $(\sigma,\nu)$ but differ in their instruments, into a single GMM system follows \citet{wooldridge2009estimating}, in which the same instrument can enter more than one equation's moment block, each instrumenting its own coefficient. For every factor $X\in\{K,L,M\}$ I impose $\mathbb{E}[D_{jt}\,\widetilde{\xi}^X_{jt}]=0$ for each component $D$ of $z^X_{jt}$: the vector of year dummies $\mathbf{d}_t$ and the eleven ownership indicators of Equation~(\ref{eq:z_def}). The year-dummy moments are the usual time-fixed-effect conditions. The ownership moments identify the ownership-type, selection, and destination-specific privatization coefficients. The lagged-input instruments are dated $t-1$. The transition and ownership-type instruments are dated $t$ but predetermined by the implementation lag. The outcome-defined indicators rest on the group mean-independence discussed there. Because exit and privatization in Chinese state-asset restructuring required an annual government inspection and an $8$--$14$ month implementation lag, as in \citet{chen2021privatization}, an Olley--Pakes-style survival correction is not required for attrition from the estimation sample. This argument covers selection into the estimation sample. The classification of the surviving state firms into kept and exiting rests on the path assumption above, with the empirical checks noted there.

\subsection{Decomposing the Error Term}

\paragraph{Identifying the Distribution of Productivity Process}

Using the parameters estimated by GMM, I isolate the residual component of each equation in system (\ref{eq:beforegmm}) by subtracting the estimated terms from both sides. This yields empirical counterparts for each factor-augmenting productivity component (multiplied by $\nu$), which also include the measurement error:

\begin{equation}
\left\{
\begin{aligned}
	& \widetilde{y^K_{jt}} = y_{jt} -\nu k_{jt} - \nu \frac{\sigma}{1-\sigma} s^K_{jt} = \nu\omega^K_{jt} + \varepsilon_{jt} \\
	& \widetilde{y^L_{jt}} = y_{jt} -\nu l_{jt} - \nu \frac{\sigma}{1-\sigma} s^L_{jt} = \nu\omega^L_{jt} + \varepsilon_{jt} \\
	& \widetilde{y^M_{jt}} = y_{jt} -\nu m_{jt} - \nu \frac{\sigma}{1-\sigma} s^M_{jt} = \nu\omega^M_{jt} + \varepsilon_{jt} \\
\end{aligned}
\right. .
\label{eq:residulas}
\end{equation}

Then the challenge is how to separate the $\omega^X_{jt}$ from the measurement error $\varepsilon_{jt}$. Recall that the productivity process $\omega^X_{jt}$ is Markovian, its innovation $\xi^X_{jt}$ is the shock, and the measurement errors $\varepsilon_{jt}$ are identically distributed across firms and over time with zero mean.\footnote{Given that the maximum time span for a single firm is 11 years, obtaining a consistent estimation of the variance for an individual firm is challenging.} I adopt a moment-matching method to identify the distribution of the productivity shocks and measurement errors.\footnote{More details of the identification of the distribution of productivity process can be found in~\ref{Appendix:B}.} This method has been used to identify the distribution of the earnings process in \citet{arellano2017earnings} and the productivity shock in \citet{aguirre2023productivity}. The joint distribution of \((\varepsilon_{j2}, ..., \varepsilon_{jT-1})\) and the joint distribution of \((\xi_{j2}^X, ..., \xi_{jT-1}^X)\) for each factor can be identified from the observations of $\widetilde{y^X_{j1}},..., \widetilde{y^X_{jT}}$.

In this setting, the variance of the measurement error ($\sigma_{\varepsilon}^2$) and the variances of the three factor-augmenting productivity shocks ($\sigma^2_{\xi^X}$) are identified by their empirical counterparts across firms. Substituting the productivity process~(\ref{eq:ar1}) into the level residual $\widetilde{y}_{t} = \nu\omega_{t} + \varepsilon_{t}$ gives the pseudo-AR(1) representation $\widetilde{y}_{t} = \rho\,\widetilde{y}_{t-1} + \nu z_{t} + \nu \xi_{t} + \varepsilon_{t} - \rho\varepsilon_{t-1}$. Taking the appropriate cross-moments, $\sigma^2_\varepsilon$ and $\sigma^2_{\xi^X}$ are recovered from
\begin{equation}
    \mathbb{E}[\widetilde{y}_{t} \ \widetilde{y}_{t-1}] = \rho \mathbb{E}[\widetilde{y}_{t-1} \widetilde{y}_{t-1} ] - \rho \sigma_{\varepsilon}^2 +\nu \mathbb{E}[z_t \widetilde{y}_{t-1}],
\end{equation}
and
\begin{equation}
    \mathbb{E}[\widetilde{y}_{t} \ \widetilde{y}_{t}] = \rho^2 \mathbb{E}[\widetilde{y}_{t-1} \widetilde{y}_{t-1} ] + (1-\rho^2)\sigma^2_{\varepsilon} + \nu^2 \sigma^2_{\xi^X} + 2\nu\rho \mathbb{E}[z_t \widetilde{y}_{t-1}] + \nu^2 \mathbb{E}[z_t^2].
\end{equation}
  Because $z_t$ is observable once the GMM coefficients in Equation~(\ref{eq:z_def}) are estimated, both moments are computable directly from the data.

\paragraph{Estimating Productivity Evolution by GLS-based Estimator}
With the two variances in hand, the three panel time sequences \(\widetilde{y}_{Xjt} = \nu\omega^X_{jt} + \varepsilon_{jt}\) are standard state-space models, and I recover the evolution of $\omega^X_{jt}$ by a Generalized Least Squares (GLS) based filter \citep{kim2017state}. The filter projects a firm's productivity in each year on its whole residual history through that year rather than on its lagged values alone, which makes it the minimum-variance linear estimator of $\omega^X_{jt}$ given information through $t$ under the AR(1) law of motion of Equation~(\ref{eq:ar1}), consistent and efficient among linear estimators even when the shocks are not normally distributed. \ref{Appendix:gls} derives it, giving the state-space form, the stacked covariance matrix, and the treatment of a firm's first observed year. Applying it to the three residual series delivers firm-year estimates $\hat\omega^K_{jt}$, $\hat\omega^L_{jt}$ and $\hat\omega^M_{jt}$, which the next section reports.

\section{Main Results} \label{sec:result}

\subsection{GMM Results}

The GMM estimates are reported in three tables corresponding to the three groups of parameters in my specification: the production-function parameters in Table~\ref{tab:gmm_results}, the ownership-type and privatization terms in Table~\ref{tab:gmm_priv}, and the year fixed effects in Table~\ref{tab:time_dummies}. I use a two-step GMM and standard errors are clustered by firm.

Table~\ref{tab:gmm_results} reports the structural parameters of the CES technology and the productivity persistence. All three factor-augmenting productivities are highly persistent: $\rho^K = 0.806$, $\rho^L = 0.842$, and $\rho^M = 0.660$. Between $66\%$ and $84\%$ of each factor's productivity carries over to the next period. Returns to scale are estimated at $\nu = 0.975$, indicating close to constant returns, and the elasticity of substitution is $\sigma = 0.384$, well below one and consistent with gross complements between the three inputs. Most of my parameter estimates are close to the past literature \citep{zhang2019non,gao2023firm}, and the few differences likely reflect differences in data, identification strategy, or my controlled-Markov productivity process, which explicitly incorporates ownership-state controls in $z^X_{jt}$ (specifications that omit these controls attribute the persistent SOE-to-private trend to AR(1) persistence and thereby change the coefficients).

\begin{table}[htbp]
\centering
\begin{threeparttable}
\caption{GMM Estimates: Production-Function Parameters.}
\label{tab:gmm_results}
\begin{tabular}{lcc}
\toprule
Parameter & Estimate & Std.\ Err.\tnote{e} \\
\midrule
$\rho^K$\tnote{a} & 0.806 & 0.001 \\
$\rho^L$\tnote{b} & 0.842 & 0.001 \\
$\rho^M$\tnote{c} & 0.660 & 0.003 \\
Elasticity of scale ($\nu$) & 0.975 & 0.001 \\
Elasticity of substitution ($\sigma$)\tnote{d} & 0.384 & 0.009 \\
\bottomrule
\end{tabular}
\begin{tablenotes}
\footnotesize
\item[a] Persistence of capital-augmenting productivity.
\item[b] Persistence of labor-augmenting productivity.
\item[c] Persistence of material-augmenting productivity.
\item[d] Standard error of $\sigma$ calculated by the delta method.
\item[e] Standard errors clustered by firm.
\item Estimated on $1{,}262{,}501$ firm-years covering $369{,}198$ firms, the analysis panel less each firm's first observed year, which is used to form the lagged instruments, and less the flicker firms of Section~\ref{sec:priv_data}.
\end{tablenotes}
\end{threeparttable}
\end{table}

\subsection{Distribution of Productivity Shock and Measurement Error}

Among the productivity innovations, the capital-augmenting shock exhibits the largest variance ($\sigma^2_{\xi^K} = 1.071$), followed by the labor-augmenting shock ($\sigma^2_{\xi^L} = 0.550$) and the material-augmenting shock ($\sigma^2_{\xi^M} = 0.067$). The relatively large $K$ and $L$ innovation variances are consistent with the substantial firm-level heterogeneity in capital and labor restructuring during the SOE-reform decade. Material-efficiency shocks are comparatively small. The measurement-error variance is estimated at $\sigma^2_\varepsilon = 0.070$.

\subsection{Estimated Factor-Augmenting Productivity}

The three factor-augmenting productivity evolutions of each firm are estimated by the GLS-based estimator.

\begin{figure}[htbp]
  \centering
  \includegraphics[width=0.8\textwidth]{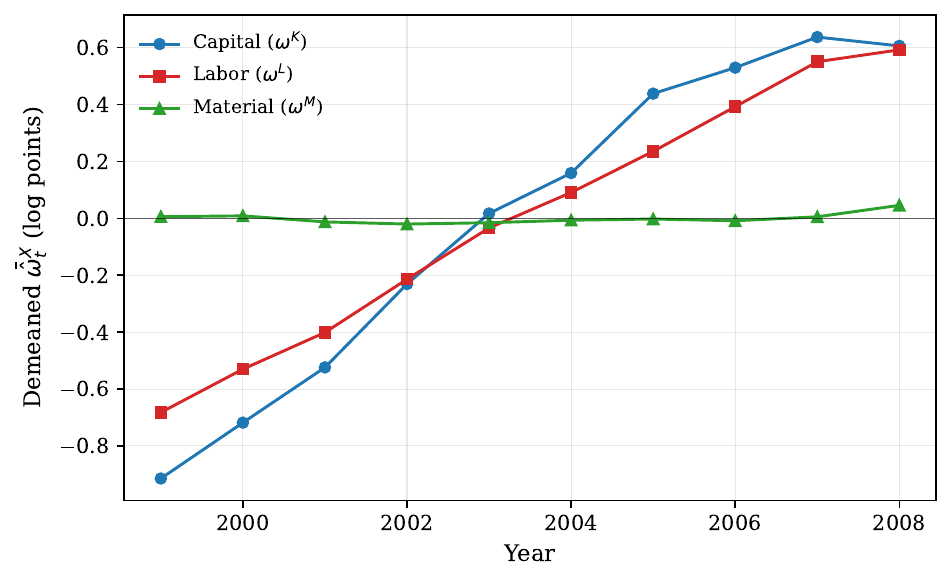}
  \caption{Year-average factor-augmenting productivity over 1999--2008, each $\hat\omega^X_t$ measured in deviations from its sample mean.}
  \label{fig:omege_3in1}
\end{figure}

Figure~\ref{fig:omege_3in1} illustrates the trends of the three factor-augmenting productivities in Chinese manufacturing, measured in differences with respect to the sample mean. Capital-augmenting productivity exhibits the most rapid increase, closely followed by labor-augmenting productivity, while material-augmenting productivity is essentially flat over the decade. The results reveal strong evidence of biased technological change with a capital-augmenting orientation.

The average annual growth rates of factor-augmenting productivity differ markedly across the three inputs.\footnote{I report the average annual growth rate as the logarithmic difference of the year-mean $\hat\omega^X_t$ between 1999 and 2008, divided by the number of years. All reported figures are unweighted firm-year averages.} Capital-augmenting productivity exhibits the highest average annual growth rate at $16.9\%$, closely followed by labor-augmenting productivity at $14.2\%$. Material-augmenting productivity is essentially flat at $0.4\%$ per year.\footnote{Because output is deflated revenue, using a two-digit industry rather than firm-specific price index, the recovered $\hat\omega^X_{jt}$ are revenue-based (TFPR, \citealp{foster2008reallocation,hsieh2009misallocation}), conflating technical efficiency with firm-specific price and markup, with $\nu$ a TFPR rather than a purely technical scale elasticity \citep{klette1996inconsistency}. \citet{chen2021privatization} note the same for these data. A common price wedge enters the three factor equations symmetrically and cancels from cross-factor comparisons such as $\hat\omega^K_{jt}-\hat\omega^L_{jt}$, largely insulating the non-neutrality of technology from markup dispersion.}

The capital-leading ranking aligns with the descriptive evidence in Section~\ref{sec:stylized}. Among the three factor cost shares, the capital share moved most over the decade, falling from $10.3\%$ to $3.8\%$, while labor and material shares moved much less. Under a CES technology with elasticity of substitution below one, a factor's cost share falls as its augmenting productivity rises relative to its price, so the sharp decline in the capital share is consistent with rapid capital-augmenting growth. Because shares also move with factor prices, however, this pattern corroborates rather than identifies the estimated ranking. The likely driver of this capital-augmenting growth is the contemporaneous SOE reform. Restructuring, together with the selective survival of large state plants, let surviving SOEs and privatizers use their capital more productively, a hypothesis I take up in Sections~\ref{sec:priv_effect} and~\ref{sec:soe_catchup}.

In the prior literature, \citet{gao2023firm} report $12.0\%$ for capital, $14.9\%$ for labor, and $0.1\%$ for materials over 1998--2007, so capital and labor lead materials as they do here, though capital sits below labor in their full-sample average. Their sub-period figures account for that. Capital grows at $25.0\%$ against labor's $12.1\%$ in 1998--2002 and then slows to $2.6\%$, so the decade average hides the early capital lead. \citet{zhang2019non} reports larger magnitudes for steel alone ($27.16\%$, $39.95\%$, and $4.80\%$).

\subsection{Ownership Productivity Gaps and the Privatization Effect} \label{sec:convergence}\label{sec:priv_effect}

\begin{figure}[htbp]
\centering
\includegraphics[width=0.8\textwidth]{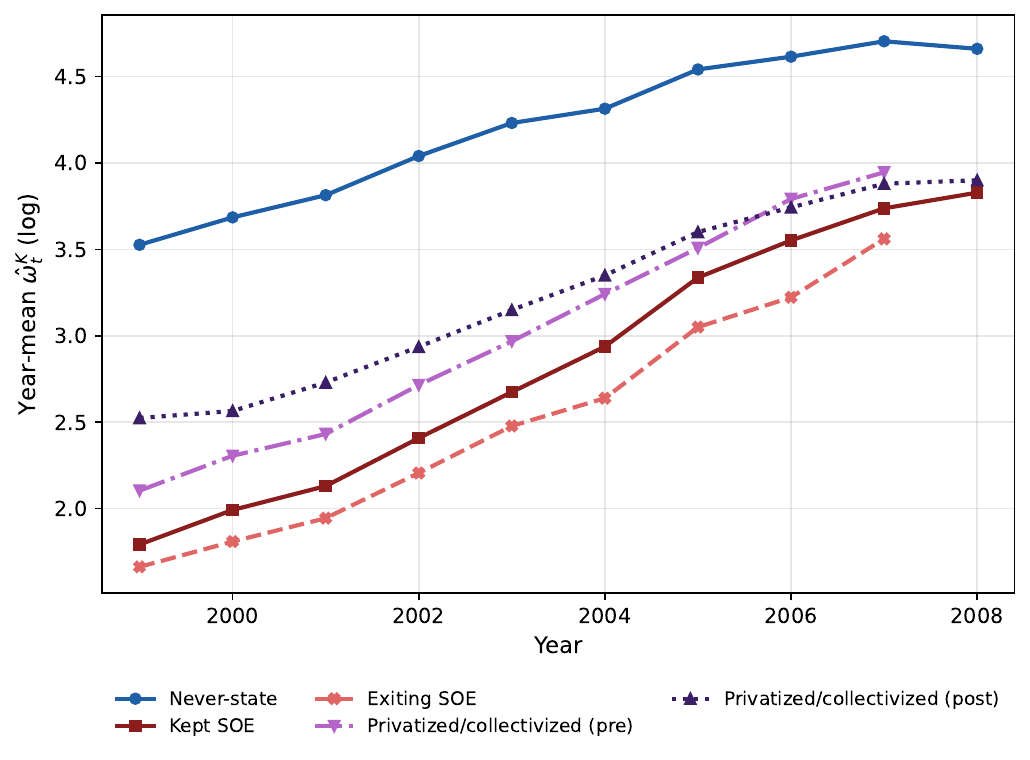}
\caption{Capital-augmenting productivity $\hat\omega^K_t$ by ownership group, 1999--2008. Groups are the five of Figure~\ref{fig:cost_shares}, which pool the two post-transition destinations and the three never-state ownership types. Their coefficients remain separate in Table~\ref{tab:gmm_priv}. The pre-privatization and exiting-SOE series end in 2007 by construction.}
\label{fig:own_k}
\end{figure}

\begin{figure}[htbp]
\centering
\includegraphics[width=0.8\textwidth]{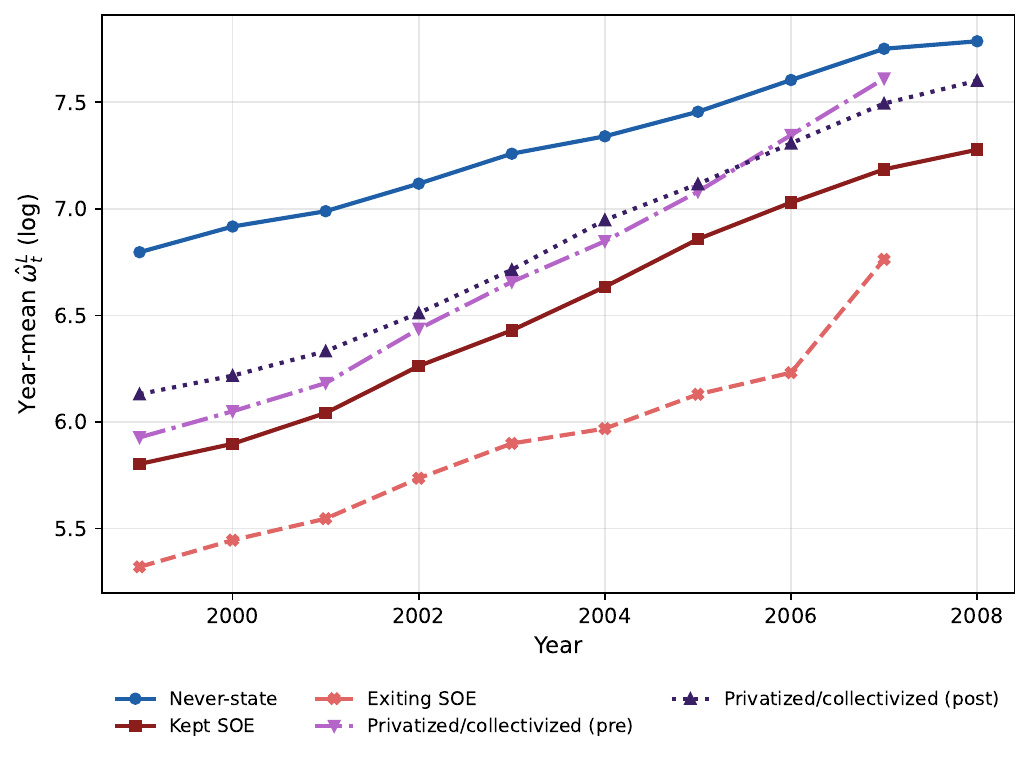}
\caption{Labor-augmenting productivity $\hat\omega^L_t$ by ownership group, 1999--2008. Groups defined as in Figure~\ref{fig:own_k}.}
\label{fig:own_l}
\end{figure}

\begin{figure}[htbp]
\centering
\includegraphics[width=0.8\textwidth]{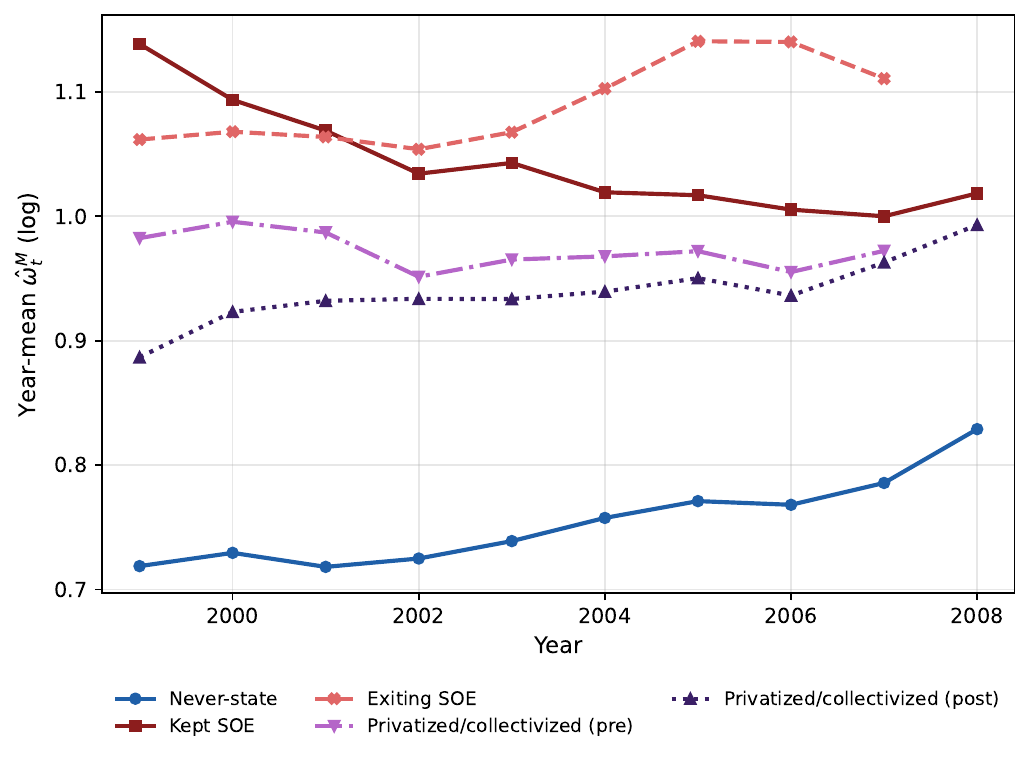}
\caption{Material-augmenting productivity $\hat\omega^M_t$ by ownership group, 1999--2008. Groups defined as in Figure~\ref{fig:own_k}.}
\label{fig:own_m}
\end{figure}

I begin by documenting how the recovered factor-augmenting productivity $\hat\omega^X_t$ evolves across the groups that the productivity process distinguishes: kept versus exiting SOEs, privatizers before and after their transition, and the never-state anchor. Figures~\ref{fig:own_k}--\ref{fig:own_m} plot the year-mean GLS-filtered $\hat\omega^X_t$ for each group. Because these groups are defined by firms' full ownership histories rather than by their current type, they hold composition roughly fixed within group, avoiding the reshuffling that contaminates a contemporaneous by-type split, in which the SOE average in 2008 is computed over a small, selected pool while the private and collective averages are diluted by inflows of former SOEs. Table~\ref{tab:gmm_priv} reports the underlying terms of the productivity process: the ownership-type terms (Panel~A), the pre-transition selection term (Panel~B), the post-transition terms by destination and timing (Panel~C), and the baseline-invariant privatization effect $\beta^{X,w}_{d}-\pi^X_{d}$ (Panel~D).

Figures~\ref{fig:own_k}--\ref{fig:own_m} make the selection and privatization margins visible. For capital (Figure~\ref{fig:own_k}), the exiting SOEs, the plants the state let go, sit well below every other group throughout the decade, while the kept SOEs climb steadily. The never-state firms form the upper frontier. Privatizers track in between: their pre-privatization path runs just above the kept SOEs, and after privatization they continue roughly parallel to the kept SOEs rather than jumping to the private frontier, matching the modest early-window privatization effect in Table~\ref{tab:gmm_priv}. Labor-augmenting productivity (Figure~\ref{fig:own_l}) shows the same ordering. The SOE-history groups, the kept SOEs and privatizers alike, catch up fastest as workforce restructuring lifts measured labor efficiency. A new finding is that material-augmenting productivity (Figure~\ref{fig:own_m}) displays the opposite ordering. The SOE groups sit at the top and the never-state firms trail throughout, with the privatized firms lying between the two. By the end of the decade the privatized firms sit at the kept-SOE level.

\begin{table}[htbp]
\centering
\renewcommand{\arraystretch}{0.9}
\begin{threeparttable}
\caption{GMM Estimates: Ownership Levels, Selection, and the Privatization Effect by Timing.}
\label{tab:gmm_priv}
\begin{tabular}{l c c c}
\toprule
 & \multicolumn{3}{c}{Factor-augmenting productivity} \\
\cmidrule(lr){2-4}
 & $\omega^{K}$ & $\omega^{L}$ & $\omega^{M}$ \\
\midrule
\multicolumn{4}{l}{\textit{Panel A: Ownership-type terms (omitted: kept-SOE firm-years)}} \\
\addlinespace
Private \ ($\alpha_P$) & $\phantom{-}0.160^{***}$ & $\phantom{-}0.103^{***}$ & $-0.081^{***}$ \\
                          & $(0.012)$ & $(0.009)$ & $(0.005)$ \\
Collective \ ($\alpha_C$) & $\phantom{-}0.174^{***}$ & $\phantom{-}0.062^{***}$ & $-0.079^{***}$ \\
                          & $(0.012)$ & $(0.009)$ & $(0.006)$ \\
Foreign \ ($\alpha_F$) & $\phantom{-}0.096^{***}$ & $\phantom{-}0.080^{***}$ & $-0.029^{***}$ \\
                          & $(0.012)$ & $(0.009)$ & $(0.005)$ \\
Exiting SOE \ ($\alpha_E$) & $-0.188^{***}$ & $-0.179^{***}$ & $\phantom{-}0.042^{***}$ \\
                          & $(0.013)$ & $(0.009)$ & $(0.006)$ \\
Other, nuisance \ ($\eta$) & $\phantom{-}0.040^{***}$ & $\phantom{-}0.051^{***}$ & $-0.027^{***}$ \\
                          & $(0.013)$ & $(0.009)$ & $(0.006)$ \\
\addlinespace
\midrule
\multicolumn{4}{l}{\textit{Panel B: Selection on trajectory, pre-privatization by destination}} \\
\addlinespace
Pre-priv., $\to$Private/Foreign \ ($\pi_{SP}$) & $\phantom{-}0.012$ & $\phantom{-}0.020^{*}$ & $\phantom{-}0.000$ \\
                          & $(0.014)$ & $(0.010)$ & $(0.006)$ \\
Pre-priv., $\to$Collective \ ($\pi_{SC}$) & $\phantom{-}0.052$ & $\phantom{-}0.035$ & $-0.044^{***}$ \\
                          & $(0.035)$ & $(0.025)$ & $(0.012)$ \\
\addlinespace
\midrule
\multicolumn{4}{l}{\textit{Panel C: Post-transition terms vs.\ kept SOEs, by destination and timing}} \\
\addlinespace
$\to$Priv./For., early \ ($\beta_{SP}^{\text{ea}}$) & $\phantom{-}0.033^{**}$ & $\phantom{-}0.043^{***}$ & $-0.018^{***}$ \\
                          & $(0.014)$ & $(0.010)$ & $(0.006)$ \\
$\to$Priv./For., late \ ($\beta_{SP}^{\text{la}}$) & $-0.037^{***}$ & $\phantom{-}0.009$ & $-0.017^{***}$ \\
                          & $(0.014)$ & $(0.010)$ & $(0.006)$ \\
$\to$Collective, early \ ($\beta_{SC}^{\text{ea}}$) & $\phantom{-}0.021$ & $-0.036^{**}$ & $-0.031^{***}$ \\
                          & $(0.026)$ & $(0.018)$ & $(0.009)$ \\
$\to$Collective, late \ ($\beta_{SC}^{\text{la}}$) & $\phantom{-}0.045$ & $-0.038$ & $-0.047^{***}$ \\
                          & $(0.035)$ & $(0.024)$ & $(0.010)$ \\
\addlinespace
\midrule
\multicolumn{4}{l}{\textit{Panel D: Baseline-invariant privatization effect $\beta_{d}^{w}-\pi_{d}$}} \\
\addlinespace
$\to$Priv./For., early & $\phantom{-}0.021^{*}$ & $\phantom{-}0.023^{***}$ & $-0.017^{***}$ \\
                          & $(0.012)$ & $(0.008)$ & $(0.004)$ \\
$\to$Priv./For., late & $-0.049^{***}$ & $-0.011$ & $-0.016^{***}$ \\
                          & $(0.012)$ & $(0.008)$ & $(0.005)$ \\
$\to$Collective, early & $-0.031$ & $-0.071^{**}$ & $\phantom{-}0.012$ \\
                          & $(0.039)$ & $(0.028)$ & $(0.013)$ \\
$\to$Collective, late & $-0.007$ & $-0.073^{**}$ & $-0.003$ \\
                          & $(0.045)$ & $(0.032)$ & $(0.014)$ \\
\bottomrule
\end{tabular}
\begin{tablenotes}
\footnotesize
\item \textit{Notes:} Early ($\text{ea}$) is the first three post-transition years ($s\in\{0,1,2\}$), late ($\text{la}$) is $s\geq3$. Standard errors clustered by firm, in parentheses (joint-covariance for the Panel-D wedges). ${}^{*}\,p<0.10$, ${}^{**}\,p<0.05$, ${}^{***}\,p<0.01$. Estimated on $1{,}262{,}501$ firm-years covering $369{,}198$ firms, the analysis panel less each firm's first observed year, which is used to form the lagged instruments, and less the flicker firms of Section~\ref{sec:priv_data}.
\end{tablenotes}
\end{threeparttable}
\end{table}

\paragraph{Ownership differences in expected productivity.}
The trajectories above are summarized structurally by the ownership-type terms in Panel~A of Table~\ref{tab:gmm_priv}, the per-period shifts in expected productivity of each group relative to the kept-SOE baseline. For capital and labor, never-SOE firms carry positive terms: $\alpha^K_P=+0.160$ for private firms, $\alpha^K_C=+0.174$ for collectives, and $\alpha^K_F=+0.096$ for foreign firms, with $+0.06$ to $+0.10$ for labor-augmenting productivity. Compounded through the persistence of the process, the private term implies a steady-state capital-augmenting gap of $0.82$ log points ($0.160/(1-\rho^K)$), about the distance between the never-state and kept-SOE trajectories in Figure~\ref{fig:own_k} at the end of the decade. For materials, the signs reverse ($\alpha^M_P=-0.081$, $\alpha^M_C=-0.079$, $\alpha^M_F=-0.029$): SOEs in my sample have systematically higher material-augmenting productivity than the non-SOE groups. This reversal likely reflects state firms' preferential access to high-quality intermediate inputs, through upstream state suppliers or administrative allocation, not superior technical efficiency. The SOEs the state allowed to fail were far below the SOEs it kept on capital and labor ($\alpha^K_E=-0.188$, $\alpha^L_E=-0.179$), the factors on which the reform's selection ran. That is direct structural evidence of ``grasp the large, let go of the small,'' and the reason the baseline isolates the kept SOEs rather than pooling them with the exiters.\footnote{Because exit responds to the productivity path itself, $\alpha^X_E$ blends a persistent type gap with the selection of negatively shocked firms into exit. I report it as a selection control rather than a causal effect of impending exit, and none of the privatization wedges rest on it (Section~\ref{sec:Estimation}).}

Panels~B, C, and~D turn from these persistent differences to the dynamics of privatization. The object of interest is the Panel~D wedge $\beta^{X,w}_{d}-\pi^X_{d}$, the shift in the conditional expectation of period-$t$ factor-augmenting productivity attributable to the transition out of state ownership in window $w$, measured against the kept SOEs and net of the destination-matched pre-privatization differential $\pi^X_d$, so that selection of SOEs into privatization is held fixed.

\paragraph{Selection on trajectory.}
Against the clean kept-SOE baseline the pre-privatization term is small for the dominant private/foreign destination: $\pi^K_{SP}=+0.012$ (not significant) and $\pi^L_{SP}=+0.020$ ($t=1.9$). Before they privatize, future privatizers are statistically indistinguishable from the kept SOEs on capital and run modestly above them on labor. The kept SOEs are therefore the right counterfactual for privatization. Before they leave state ownership, the privatizers resemble the SOEs the state kept ($\pi^X$ near zero), not the far less productive plants it let go ($\alpha^X_E$ large and negative), so the kept SOEs, rather than a pooled-SOE average, are the group these firms would have belonged to had they stayed in state hands.

\paragraph{Privatization raises labor-augmenting productivity at the transition.}
Splitting the post-transition term by timing shows what a single pooled coefficient averages away. For the dominant flow to private or foreign ownership, the labor-augmenting privatization effect is positive and concentrated in the early years out of state ownership: the early wedge is $\beta^{L,\mathrm{ea}}_{SP}-\pi^L_{SP}=+0.023$, fading to $-0.011$ in the late window, an early-minus-late contrast of $+0.034$ that identifies the front-loading directly. Privatizing firms shed the administratively inflated employment that soft-budget guarantees had sustained, raising measured labor-augmenting productivity in the transition window, after which they improve no faster than the kept SOEs. Capital-augmenting productivity shows the mirror image of a transitory adjustment: a marginal early gain ($\beta^{K,\mathrm{ea}}_{SP}-\pi^K_{SP}=+0.021$) is followed by a significant late decline ($-0.049$), leaving privatized firms slightly below the kept SOEs on capital by the end of the decade, consistent with those plants' own rapid capital-augmenting growth (Section~\ref{sec:soe_catchup}). Material-augmenting productivity falls modestly and steadily in both windows. The collective destination carries a labor penalty throughout ($\beta^{L,\mathrm{ea}}_{SC}-\pi^L_{SC}=-0.071$ and $\beta^{L,\mathrm{la}}_{SC}-\pi^L_{SC}=-0.073$), so the two destinations differ in how labor absorbs the transition.

The pooled single-coefficient specification of the previous literature averages this timing to near zero. With one post-transition term per destination (Appendix Table~\ref{tab:gmm_priv_pooled}) the labor effect is $\beta^L_{SP}-\pi^L=+0.008$, the mean of a significant early gain and a null late window. An event study of the recovered $\hat\omega^X$ against the kept SOEs traces the same path directly. Labor-augmenting productivity is flat before the transition, rises over the first two to three years, and holds that higher level thereafter, while capital rises briefly and then falls below the kept SOEs. \ref{Appendix:eventstudy} reports the estimates and the tests for pre-trends.

\subsection{Privatization Effects on the 1998 Cohort} \label{sec:cohort98}

The full-panel analysis of Section~\ref{sec:priv_effect} sits against a literature that
locates the productivity gains of privatization in a specific population: the incumbent
state firms that were the actual targets of the late-1990s restructuring.
\citet{chen2021privatization} accordingly restrict their main analysis to the $1998$
cohort (firms already on record at the start of the sample) rather than the full panel,
whose later years fill with newly entering SOE firms that have less overstaffing and so carry less privatization treatment. I replicate that restriction within my
framework. The $1998$ cohort comprises $73{,}728$ firms and $379{,}673$ firm-years, and I
re-estimate the main specification on this
subsample. Table~\ref{tab:cohort98} reports the estimates. The ownership-type terms are larger within this incumbent cohort than on the full panel ($\alpha^K_P=+0.217$ and $\alpha^K_E=-0.260$, against $+0.160$ and $-0.188$ in Table~\ref{tab:gmm_priv}), and with the cohort's higher persistence ($\rho^K=0.862$) the implied steady-state gaps roughly double. The incumbents among which the reform sorted were more dispersed in factor-augmenting productivity than the panel at large.

\begin{table}[htbp]
\centering
\renewcommand{\arraystretch}{0.9}
\begin{threeparttable}
\caption{Privatization Effect on the 1998 Cohort, by Timing.}
\label{tab:cohort98}
\begin{tabular}{lccc}
\toprule
 & $\omega^K$ & $\omega^L$ & $\omega^M$ \\
\midrule
\multicolumn{4}{l}{\textit{Panel A: Ownership-type terms (omitted: kept-SOE)}} \\
\addlinespace
Private \ ($\alpha_P$) & $\phantom{-}0.217^{***}$ & $\phantom{-}0.218^{***}$ & $-0.150^{***}$ \\
  & $(0.028)$ & $(0.027)$ & $(0.015)$ \\
Collective \ ($\alpha_C$) & $\phantom{-}0.261^{***}$ & $\phantom{-}0.219^{***}$ & $-0.158^{***}$ \\
  & $(0.030)$ & $(0.028)$ & $(0.015)$ \\
Foreign \ ($\alpha_F$) & $\phantom{-}0.190^{***}$ & $\phantom{-}0.186^{***}$ & $-0.078^{***}$ \\
  & $(0.028)$ & $(0.025)$ & $(0.013)$ \\
Exiting SOE \ ($\alpha_E$) & $-0.260^{***}$ & $-0.238^{***}$ & $\phantom{-}0.075^{***}$ \\
  & $(0.028)$ & $(0.024)$ & $(0.014)$ \\
Other, nuisance \ ($\eta$)\tnote{$\dagger$} & $\phantom{-}0.094^{***}$ & $\phantom{-}0.108^{***}$ & $-0.077^{***}$ \\
  & $(0.028)$ & $(0.025)$ & $(0.014)$ \\
\addlinespace
\midrule
\multicolumn{4}{l}{\textit{Panel B: Selection, pre-privatization by destination}} \\
\addlinespace
Pre-priv., $\to$Priv./For. \ ($\pi_{SP}$) & $\phantom{-}0.059^{**}$ & $\phantom{-}0.011$ & $-0.020$ \\
  & $(0.028)$ & $(0.024)$ & $(0.013)$ \\
Pre-priv., $\to$Collective \ ($\pi_{SC}$) & $\phantom{-}0.128^{**}$ & $\phantom{-}0.056$ & $-0.071^{***}$ \\
  & $(0.065)$ & $(0.051)$ & $(0.024)$ \\
\addlinespace
\midrule
\multicolumn{4}{l}{\textit{Panel C: Post-transition terms vs.\ kept SOEs, by destination and timing}} \\
\addlinespace
$\to$Priv./For., early \ ($\beta_{SP}^{\text{ea}}$) & $\phantom{-}0.104^{***}$ & $\phantom{-}0.092^{***}$ & $-0.051^{***}$ \\
  & $(0.030)$ & $(0.025)$ & $(0.014)$ \\
$\to$Priv./For., late \ ($\beta_{SP}^{\text{la}}$) & $\phantom{-}0.036$ & $\phantom{-}0.123^{***}$ & $-0.075^{***}$ \\
  & $(0.029)$ & $(0.027)$ & $(0.014)$ \\
$\to$Collective, early \ ($\beta_{SC}^{\text{ea}}$) & $-0.012$ & $\phantom{-}0.021$ & $-0.058^{***}$ \\
  & $(0.053)$ & $(0.043)$ & $(0.020)$ \\
$\to$Collective, late \ ($\beta_{SC}^{\text{la}}$) & $\phantom{-}0.115^{*}$ & $\phantom{-}0.032$ & $-0.108^{***}$ \\
  & $(0.063)$ & $(0.052)$ & $(0.020)$ \\
\addlinespace
\midrule
\multicolumn{4}{l}{\textit{Panel D: Baseline-invariant effect $\beta_{d}^{w}-\pi_{d}$}} \\
\addlinespace
$\to$Priv./For., early & $\phantom{-}0.045^{*}$ & $\phantom{-}0.081^{***}$ & $-0.032^{***}$ \\
  & $(0.025)$ & $(0.018)$ & $(0.009)$ \\
$\to$Priv./For., late & $-0.023$ & $\phantom{-}0.112^{***}$ & $-0.055^{***}$ \\
  & $(0.023)$ & $(0.020)$ & $(0.009)$ \\
$\to$Collective, early & $-0.140^{*}$ & $-0.035$ & $\phantom{-}0.013$ \\
  & $(0.075)$ & $(0.057)$ & $(0.025)$ \\
$\to$Collective, late & $-0.013$ & $-0.024$ & $-0.037$ \\
  & $(0.082)$ & $(0.067)$ & $(0.025)$ \\
\bottomrule
\end{tabular}
\begin{tablenotes}
\footnotesize
\item \textit{Notes:} The main specification (destination-split pre-privatization terms and early/late post-transition terms) re-estimated on the $1998$ cohort (firms on record in $1998$; $73{,}728$ firms, $379{,}673$ firm-years). Early is $s\in\{0,1,2\}$, late is $s\geq3$. The cohort returns $\hat\sigma=0.68$, $\hat\nu=0.92$, $\rho=(0.862,0.85,0.71)$; steady-state gaps multiply by $1/(1-\rho^X)$ ($7.2$, $6.7$, $3.4$). Standard errors clustered by firm (joint-covariance for Panel-D wedges).
\item[$\dagger$] $\eta$ estimated, not interpreted.
\item ${}^{*}\,p<0.10$, ${}^{**}\,p<0.05$, ${}^{***}\,p<0.01$.
\end{tablenotes}
\end{threeparttable}
\end{table}

Restricting attention to the incumbent cohort recovers a privatization effect on
labor-augmenting productivity that is both larger and more durable than on the full panel. For
the dominant transition to private or foreign ownership, the labor gain is significant in both
timing windows and, unlike the early-window full-panel effect, does not fade: the early wedge is
$\beta^{L,\mathrm{ea}}_{SP}-\pi^L_{SP}=+0.081$ and the late wedge is
$\beta^{L,\mathrm{la}}_{SP}-\pi^L_{SP}=+0.112$. Capital shows no comparable gain (a marginal early
$+0.045$, and an insignificant late $-0.023$) and material-augmenting productivity
falls in both windows ($-0.032$ and $-0.055$). As on the full panel (Section~\ref{sec:priv_effect}), privatization lets these firms shed surplus labor. The $1998$ incumbents carry the heaviest legacy overstaffing, so the restructuring continues over several years and the gain persists into the late window, while the full panel's post-$1998$ SOE entrants carry less of that legacy and pull the average down. The effect
is thus concentrated in the same incumbent cohort that \citet{chen2021privatization} treat as the
reform's main target, and my factor decomposition adds that it runs through labor mostly, a finding that one TFP cannot capture.

\section{Heterogeneity across Industries} \label{sec:heterogeneity}

\subsection{Industry-Level CES Technology} \label{sec:industry_tech}

I re-estimate the full specification separately within each of ten manufacturing industries. The industry definitions are listed in Table~\ref{tab:industry_classification} and follow \citet{jaumandreu2018cost} up to two minor adjustments described in~\ref{Appendix:data}. All ten industries have $\hat\sigma$ between $0.15$ and $0.72$ and $\hat\nu$ close to one (Table~\ref{tab:industry_prod}). The factor-augmenting decomposition is therefore well defined sector by sector, not an artifact of the pooled sample. Material-augmenting productivity is the least persistent factor in every industry.

\begin{table}[htbp]
\centering
\begin{threeparttable}
\caption{Industry Heterogeneity: Production-Function Parameters.}
\label{tab:industry_prod}
{\small
\renewcommand{\arraystretch}{0.95}
\begin{tabular}{lccccc}
\toprule
Industry & $\hat\sigma$\tnote{a} & $\hat\nu$ & $\hat\rho^K$ & $\hat\rho^L$ & $\hat\rho^M$ \\
\midrule
Food, drink and tobacco & $0.392$ & $0.985$ & $0.790$ & $0.805$ & $0.672$ \\
 & $(0.024)$ & $(0.004)$ & $(0.004)$ & $(0.003)$ & $(0.010)$ \\
\addlinespace
Textile, leather and shoes & $0.152$ & $0.989$ & $0.815$ & $0.846$ & $0.635$ \\
 & $(0.020)$ & $(0.002)$ & $(0.002)$ & $(0.003)$ & $(0.010)$ \\
\addlinespace
Timber and furniture & $0.183$ & $0.987$ & $0.783$ & $0.800$ & $0.551$ \\
 & $(0.043)$ & $(0.004)$ & $(0.006)$ & $(0.007)$ & $(0.023)$ \\
\addlinespace
Paper & $0.578$ & $0.983$ & $0.833$ & $0.759$ & $0.552$ \\
 & $(0.069)$ & $(0.007)$ & $(0.006)$ & $(0.010)$ & $(0.020)$ \\
\addlinespace
Printing & $0.452$ & $0.998$ & $0.854$ & $0.838$ & $0.684$ \\
 & $(0.053)$ & $(0.010)$ & $(0.007)$ & $(0.008)$ & $(0.019)$ \\
\addlinespace
Chemical products & $0.404$ & $0.981$ & $0.816$ & $0.833$ & $0.690$ \\
 & $(0.021)$ & $(0.004)$ & $(0.002)$ & $(0.003)$ & $(0.008)$ \\
\addlinespace
Non-metallic minerals & $0.188$ & $0.972$ & $0.828$ & $0.862$ & $0.544$ \\
 & $(0.027)$ & $(0.002)$ & $(0.003)$ & $(0.004)$ & $(0.011)$ \\
\addlinespace
Metals and metal products & $0.723$ & $0.915$ & $0.773$ & $0.819$ & $0.597$ \\
 & $(0.032)$ & $(0.009)$ & $(0.004)$ & $(0.003)$ & $(0.013)$ \\
\addlinespace
Machinery and transport equipment & $0.583$ & $0.964$ & $0.791$ & $0.814$ & $0.632$ \\
 & $(0.026)$ & $(0.004)$ & $(0.003)$ & $(0.003)$ & $(0.008)$ \\
\addlinespace
Electronics & $0.563$ & $0.975$ & $0.807$ & $0.843$ & $0.692$ \\
 & $(0.029)$ & $(0.005)$ & $(0.003)$ & $(0.003)$ & $(0.010)$ \\
\bottomrule
\end{tabular}}
\begin{tablenotes}
\footnotesize
\item \textit{Notes:} Industry-by-industry estimates from the main specification; standard errors below each estimate, clustered by firm.
\item[a] Standard error of $\hat\sigma$ by the delta method.
\end{tablenotes}
\end{threeparttable}
\end{table}

\subsection{Factor-Augmenting Productivity Growth across Industries} \label{sec:sector_growth_omega}

From these per-industry estimates I filter the firm-year $\hat\omega^K_{jt}$, $\hat\omega^L_{jt}$, and $\hat\omega^M_{jt}$. Table~\ref{tab:sector_omega_growth} reports their annual average growth by industry.

\begin{table}[htbp]
\centering
\begin{threeparttable}
\caption{Annual Average Growth of Factor-Augmenting Productivity by Industry (1999--2008).}
\label{tab:sector_omega_growth}
\begin{tabular}{lccc}
\toprule
Industry & $g_{\omega^K}$ & $g_{\omega^L}$ & $g_{\omega^M}$ \\
 & (\%/yr) & (\%/yr) & (\%/yr) \\
\midrule
Food, drink and tobacco & \phantom{-}15.1 & \phantom{-}13.3 & -1.4 \\
Textile, leather and shoes & \phantom{-}10.1 & \phantom{-}10.5 & \phantom{-}1.0 \\
Timber and furniture & \phantom{-}11.8 & \phantom{-}12.2 & \phantom{-}0.6 \\
Paper & \phantom{-}25.5 & \phantom{-}17.1 & -0.5 \\
Printing & \phantom{-}19.1 & \phantom{-}21.6 & -2.1 \\
Chemical products & \phantom{-}15.6 & \phantom{-}14.1 & \phantom{-}0.8 \\
Non-metallic minerals & \phantom{-}14.7 & \phantom{-}18.7 & \phantom{-}1.3 \\
Metals and metal products & \phantom{-}34.2 & \phantom{-}14.2 & -0.3 \\
Machinery and transport equipment & \phantom{-}28.1 & \phantom{-}20.4 & -1.0 \\
Electronics & \phantom{-}25.4 & \phantom{-}12.3 & \phantom{-}1.2 \\
\bottomrule
\end{tabular}
\begin{tablenotes}
\footnotesize
\item \textit{Notes:} Average annual growth of the filter-recovered factor-augmenting productivities $\hat\omega^X$, computed as the change in the unweighted year-mean $\hat\omega^X$ between $1999$ and $2008$ divided by the number of years, in log points per year.
\end{tablenotes}
\end{threeparttable}
\end{table}

Capital-augmenting productivity grew the fastest in most industries and is also the input whose growth differs most across them. Capital-augmenting growth ranges from about $10\%$ per year in textiles to above $30\%$ in metals and close to $30\%$ in machinery and transport equipment, a much wider spread than for labor, while material-augmenting productivity stays close to zero everywhere.

Technical change is also more capital-biased in several of the heavy industries than in the light ones. The gap between capital- and labor-augmenting growth is large and positive in metals, electronics, paper, and machinery and transport equipment, heavy industries that entered the reform with large stocks of aging equipment, and close to zero or slightly labor-biased in lighter industries such as textiles/leather/shoes, timber/furniture, and printing. The main exception is non-metallic minerals, a heavy industry in which labor-augmenting growth outpaces capital ($18.7\%$ against $14.7\%$ per year), plausibly reflecting the labor restructuring of cement and glass plants.

\subsection{The Privatization Effect across Industries} \label{sec:industry_priv}

The per-industry runs of Section~\ref{sec:industry_tech} (kept-SOE baseline, clean treatment labels) also deliver the privatization effects. Table~\ref{tab:industry} reports, for each industry, the baseline-invariant privatization effect $\beta^{X,w}_{d}-\pi^X_d$ by destination (SOE$\to$Private/Foreign and SOE$\to$Collective) and timing window (early and late). The full set of per-industry coefficients is reported in \ref{Appendix:industry_full}. The coefficients are per-period terms of each industry's productivity process, so cross-industry comparisons of implied long-run effects also reflect differences in the industry persistences of Table~\ref{tab:industry_prod}.

\begin{table}[htbp]
\centering
\begin{threeparttable}
\caption{Industry Heterogeneity: Privatization Effect $\beta_{d}^{w}-\pi_{d}$ by Destination and Timing.}
\label{tab:industry}
{\small
\renewcommand{\arraystretch}{0.95}
\setlength{\tabcolsep}{5pt}
\begin{tabular}{lcccc}
\toprule
 & \multicolumn{2}{c}{SOE$\to$Private/Foreign} & \multicolumn{2}{c}{SOE$\to$Collective} \\
\cmidrule(lr){2-3} \cmidrule(lr){4-5}
Industry & early & late & early & late \\
\midrule
\multicolumn{5}{l}{\textit{Panel A: Capital-augmenting efficiency $\omega^K$}} \\
Food, drink and tobacco & $-0.001$ & $-0.075^{**}$ & $-0.288^{***}$ & $-0.211^{*}$ \\
Textile, leather and shoes & $\phantom{-}0.031$ & $-0.052$ & $-0.134$ & $-0.184^{*}$ \\
Timber and furniture & $-0.033$ & $-0.014$ & $\phantom{-}0.381^{\dagger}$ & $\phantom{-}0.342^{\dagger}$ \\
Paper & $\phantom{-}0.081$ & $\phantom{-}0.012$ & $\phantom{-}0.241^{**\dagger}$ & $\phantom{-}0.048^{\dagger}$ \\
Printing & $-0.038$ & $-0.064$ & $-0.175^{\dagger}$ & $-0.354^{\dagger}$ \\
Chemical products & $\phantom{-}0.010$ & $-0.114^{***}$ & $\phantom{-}0.126$ & $\phantom{-}0.108$ \\
Non-metallic minerals & $-0.019$ & $-0.107^{***}$ & $\phantom{-}0.050$ & $\phantom{-}0.060$ \\
Metals and metal products & $\phantom{-}0.220^{**}$ & $\phantom{-}0.113$ & $\phantom{-}0.038$ & $-0.209$ \\
Machinery and transport equipment & $\phantom{-}0.063$ & $\phantom{-}0.037$ & $-0.219$ & $-0.031$ \\
Electronics & $\phantom{-}0.010$ & $-0.032$ & $\phantom{-}0.095$ & $\phantom{-}0.267^{*}$ \\
\midrule
\multicolumn{5}{l}{\textit{Panel B: Labor-augmenting efficiency $\omega^L$}} \\
Food, drink and tobacco & $\phantom{-}0.020$ & $-0.046^{*}$ & $-0.227^{***}$ & $-0.178^{*}$ \\
Textile, leather and shoes & $\phantom{-}0.011$ & $-0.034$ & $-0.145^{**}$ & $-0.176^{**}$ \\
Timber and furniture & $-0.034$ & $-0.000$ & $\phantom{-}0.006^{\dagger}$ & $\phantom{-}0.112^{\dagger}$ \\
Paper & $-0.032$ & $-0.039$ & $\phantom{-}0.324^{**\dagger}$ & $\phantom{-}0.384^{*\dagger}$ \\
Printing & $\phantom{-}0.038$ & $\phantom{-}0.020$ & $-0.069^{\dagger}$ & $\phantom{-}0.068^{\dagger}$ \\
Chemical products & $\phantom{-}0.058^{***}$ & $\phantom{-}0.005$ & $\phantom{-}0.002$ & $-0.021$ \\
Non-metallic minerals & $-0.009$ & $-0.064^{***}$ & $-0.027$ & $-0.095$ \\
Metals and metal products & $\phantom{-}0.012$ & $\phantom{-}0.126$ & $-0.162$ & $-0.614$ \\
Machinery and transport equipment & $\phantom{-}0.064^{**}$ & $\phantom{-}0.057^{**}$ & $-0.078$ & $-0.046$ \\
Electronics & $\phantom{-}0.028$ & $\phantom{-}0.045$ & $\phantom{-}0.139$ & $\phantom{-}0.312^{***}$ \\
\midrule
\multicolumn{5}{l}{\textit{Panel C: Material-augmenting efficiency $\omega^M$}} \\
Food, drink and tobacco & $-0.015$ & $\phantom{-}0.001$ & $\phantom{-}0.081^{***}$ & $\phantom{-}0.054^{*}$ \\
Textile, leather and shoes & $-0.010$ & $-0.027^{**}$ & $\phantom{-}0.031$ & $\phantom{-}0.056^{**}$ \\
Timber and furniture & $\phantom{-}0.002$ & $-0.034$ & $-0.034^{\dagger}$ & $-0.106^{**\dagger}$ \\
Paper & $\phantom{-}0.020$ & $-0.051^{*}$ & $-0.114^{***\dagger}$ & $-0.151^{**\dagger}$ \\
Printing & $\phantom{-}0.008$ & $\phantom{-}0.004$ & $\phantom{-}0.126^{\dagger}$ & $\phantom{-}0.101^{\dagger}$ \\
Chemical products & $-0.005$ & $\phantom{-}0.025^{**}$ & $-0.007$ & $\phantom{-}0.002$ \\
Non-metallic minerals & $-0.029^{***}$ & $-0.046^{***}$ & $-0.019$ & $-0.033$ \\
Metals and metal products & $-0.046^{*}$ & $-0.105^{***}$ & $\phantom{-}0.070$ & $\phantom{-}0.132$ \\
Machinery and transport equipment & $-0.046^{***}$ & $-0.076^{***}$ & $\phantom{-}0.019$ & $-0.017$ \\
Electronics & $-0.001$ & $-0.009$ & $-0.067$ & $-0.183^{***}$ \\
\bottomrule
\end{tabular}}
\begin{tablenotes}
\footnotesize
\item \textit{Notes:} Per-industry estimates, the industry counterpart of Panel~D of the pooled Table~\ref{tab:gmm_priv}. Each cell is the baseline-invariant privatization effect $\beta_{d}^{w}-\pi_{d}$ for the early ($s\in\{0,1,2\}$) and late ($s\geq3$) windows, with standard error from the joint parameter covariance.
\item[$\dagger$] SOE$\to$Collective cells (early and late) in industries with fewer than $100$ early firm-years (imprecise).
\item Standard errors clustered by firm. ${}^{*}\,p<0.10$, ${}^{**}\,p<0.05$, ${}^{***}\,p<0.01$.
\end{tablenotes}
\end{threeparttable}
\end{table}

\paragraph{Patterns across industries.}
For capital, the early effect is close to zero in most industries. The exception is metals and metal products, where the early effect is significantly positive. The late capital effect turns significantly negative in chemicals and non-metallic minerals, and in food, drink and tobacco, the industry analogue of the pooled early-gain-then-late-decline capital path. For labor, the early effect is significantly positive where administrative overstaffing was heaviest, chemicals and machinery and transport. In machinery the labor gain persists into the late window, as in the $1998$ cohort, while in the late window it turns significantly negative in non-metallic minerals and food. For materials, the effect is negative in both windows across metals, machinery and transport, and non-metallic minerals, consistent with the pooled material decline. The collective destination is too noisy to read in the light sectors with thin early cells (timber, paper, and printing). Where it is precisely estimated it does not consistently follow the private channel, so I treat the private and foreign destination as the cleaner privatization margin.

\paragraph{Interpretation.}
Net of the destination-matched pre-privatization selection $\pi^X_d$ and against the kept-SOE baseline, the single significant positive early capital effect, in metals, arises where the reform's exit selection ran deepest ($\alpha^K_E=-0.52$, \ref{Appendix:industry_full}) and where the privatizers themselves start lowest relative to the kept SOEs, with the most negative pre-privatization term of the ten industries ($\pi^K_{SP}=-0.10$, imprecisely estimated). They close part of that gap at the transition.\footnote{Like the pooled early capital wedge, this estimate is an upper bound under an SOE capital subsidy (Section~\ref{sec:capwedge}).} The negative material-augmenting effects across heavy industry fit the input-access channel. State firms enjoy preferential access to intermediate inputs (Section~\ref{sec:result}), which privatized plants lose, and their material efficiency falls. The pattern also fits \citet{chen2021privatization}, who attribute negative privatization effects to the short-run adjustment costs of the organizational disruption that accompanies the ownership change, largest for the bigger SOEs concentrated in these sectors.

\section{Discussion: Privatization versus Survival in the SOE Catch-Up} \label{sec:discussion}\label{sec:soe_catchup}

The modest, front-loaded privatization effect of Section~\ref{sec:priv_effect} prompts a broader question: state-owned firms clearly close their productivity gap with non-SOEs over the reform decade, but through which margin? Three forces operate at once: the improvement of firms that remain state-owned, the selective exit of low-productivity state plants, and the durable privatization of others.

\paragraph{The catch-up is real.}
Figures~\ref{fig:own_k}--\ref{fig:own_m} show the firms the reform touched closing much of their productivity gap with never-state firms over the decade: those that stay state-owned close roughly half of both the capital gap and the labor gap against the non-state benchmark, while the durable privatizers close about a quarter of the capital gap and some seventy percent of the labor gap. This aggregate catch-up, however, runs together three forces that the year-mean trajectories cannot separate: the entry of state firms into the sample, the exit of others, and the improvement of the plants that stay state-owned.

\paragraph{A formal within--selection decomposition.}
To quantify these channels I apply the \citet{melitz2015dynamic} dynamic Olley--Pakes decomposition to the change in the log-sales-weighted mean factor-augmenting productivity of the state sector between $1999$ and $2008$. Survivors ($S$) are state-owned in both years. Exiters ($X$) are state-owned in $1999$ but not $2008$, split into firms that leave the panel ($D$) and firms that remain but are no longer state-owned ($P$), with $X=D\cup P$. Entrants ($E$) are state-owned in $2008$ but not $1999$.

Let $\mu_{jt}=\log\text{sales}_{jt}\big/\!\sum_{k\in\mathrm{SOE}_t}\log\text{sales}_{kt}$ be firm $j$'s share of log sales within the year-$t$ state sector,\footnote{I use the log of sales rather than the raw sales level to limit the leverage of a few very large state plants on the reallocation terms. \citet{doraszelski2018measuring} adopt the same share-of-log-sales weight in their Olley--Pakes/Melitz--Polanec decomposition of factor-augmenting productivity.} $\mu^G_t=\sum_{j\in G}\mu_{jt}$ the weight of group $G$, $\Phi^{X}_{G,t}=\sum_{j\in G}(\mu_{jt}/\mu^G_t)\,\hat\omega^X_{jt}$ its log-sales-weighted mean, and $\bar\omega^{X,S}$ the survivors' unweighted mean over their $N^S$ firms. With $t_1\equiv1999$, $t_2\equiv2008$, the change decomposes exactly as
\begin{equation}\label{eq:mp}
\begin{aligned}
\Delta\Phi^X_{\mathrm{SOE}}
={}& \underbrace{\Delta\bar\omega^{X,S} + N^S\,\Delta\mathrm{Cov}\!\big(\mu/\mu^S,\hat\omega^X\big)}_{\text{survivors: shift + reallocation}}
+ \underbrace{\mu^E_{t_2}\big(\Phi^{X}_{E,t_2}-\Phi^{X}_{S,t_2}\big)}_{\text{entry}} \\
&+ \underbrace{\mu^D_{t_1}\big(\Phi^{X}_{S,t_1}-\Phi^{X}_{D,t_1}\big)}_{\text{exit: death}}
+ \underbrace{\mu^P_{t_1}\big(\Phi^{X}_{S,t_1}-\Phi^{X}_{P,t_1}\big)}_{\text{exit: privatized out}},
\end{aligned}
\end{equation}
The reallocation term is $N^S$ times the change in the covariance between a survivor's within-group share $\mu_{jt}/\mu^S_t$ and its productivity. Table~\ref{tab:soe_mp} reports the result.

The state sector's mean productivity rises sharply over the decade, by $2.04$ log points for $\hat\omega^K$ and $1.96$ for $\hat\omega^L$, and Table~\ref{tab:soe_mp} traces the sources. For both factors the rise comes mostly from the within-survivor distribution shift (about three-quarters for capital and $61\%$ for labor) and from the entry of state firms into the sample, which adds a further $28\%$ for capital and $26\%$ for labor. Entry, whether of newly observed or newly state-owned plants, is a composition margin rather than a treatment of the incumbent state plants, so the discussion below centers on the within and exit margins. The two factors differ in the exit margin: for capital, reallocation and selection are small (selection slightly negative), whereas for labor the exit of overstaffed plants contributes $11\%$, almost entirely from low-labor-productivity firms leaving the panel rather than from those privatized while remaining in it.

\begin{table}[htbp]\centering
\begin{threeparttable}
\caption{Melitz--Polanec Decomposition of the Change in the Factor-Augmenting Productivity of the State Sector, 1999--2008.}
\label{tab:soe_mp}
\begin{tabular}{lccc}
\toprule
 & $\hat\omega^K$ & $\hat\omega^L$ & $\hat\omega^M$ \\
\midrule
Total change $\Delta\Phi_{\text{SOE}}$ & $+2.043$ & $+1.958$ & $-0.024$ \\
\midrule
\quad Within survivors (distribution shift) & $+1.573$ & $+1.192$ & $-0.031$ \\
\quad Reallocation (covariance) & $+0.006$ & $+0.036$ & $-0.007$ \\
\quad Exit / selection & $-0.111$ & $+0.213$ & $+0.121$ \\
\qquad of which: panel exit (death) & $-0.064$ & $+0.244$ & $+0.104$ \\
\qquad of which: privatized out & $-0.047$ & $-0.031$ & $+0.017$ \\
\quad Entry & $+0.574$ & $+0.517$ & $-0.107$ \\
\bottomrule
\end{tabular}
\begin{tablenotes}\footnotesize
\item \textit{Notes:} Survivors are in both pools; exiters are in the 1999 pool only, split into firms that leave the panel by 2008 ($D$) and firms still in the panel but no longer state-owned ($P$); entrants are in the 2008 pool only.
\end{tablenotes}
\end{threeparttable}
\end{table}

\paragraph{Why the privatization wedge is near zero.}
If kept SOEs improve so much, why do privatizers, measured against those same kept SOEs, show no gain? Table~\ref{tab:soe_within_growth} answers it directly by comparing growth rates across ownership groups in the balanced panel, where composition is fixed. First, the near-zero wedge does not mean privatized firms stagnated. They improve rapidly in absolute terms: on capital no faster than the SOEs the state kept ($16.8$ against $17.5\%$ a year), and on labor about two points a year faster ($15.0$ against $13.2$), matching the front-loaded transition gain of Section~\ref{sec:priv_effect} and the persistent level gain in the event study (\ref{Appendix:eventstudy}). The capital gains they display would have come without the ownership change, and so would the labor growth beyond the restructuring step of the transition window. Second, no choice of counterfactual group revives the effect. Folding the exiting SOEs back into the omitted baseline leaves the wedge nearly unchanged (\ref{Appendix:baseline}). Privatized firms do not take on the always-private type's productivity profile either: their post-transition terms settle well below the never-state private terms on capital and labor (Table~\ref{tab:gmm_priv}). A current-type specification such as \citet{chen2021privatization}'s imposes that convergence by construction. Mine leaves the post-transition terms free. The early-window labor gain of Section~\ref{sec:priv_effect} is larger and persistent within the incumbent cohort (Section~\ref{sec:cohort98}) and in the most overstaffed industries (Section~\ref{sec:industry_priv}). Whichever reference group one takes, the ownership change does not transform a firm's measured factor efficiency.

\begin{table}[htbp]\centering
\begin{threeparttable}
\caption{Within-Firm Annual Growth by Ownership Group, Balanced Panel of Firms Present Every Year 1999--2008 (\% per Year).}
\label{tab:soe_within_growth}
\small
\setlength{\tabcolsep}{6pt}
\begin{tabular}{l ccc c cccc}
\toprule
Group & $\hat\omega^K$ & $\hat\omega^L$ & $\hat\omega^M$ & & $y$ & $k$ & $l$ & $m$ \\
\midrule
Always-private     & 10.5 & 11.0 & \phantom{-}1.6 & & 13.6 & \phantom{-}8.0 & \phantom{-}3.0 & 12.1 \\
Always-collective  & 13.9 & 12.3 & \phantom{-}1.5 & & 12.1 & \phantom{-}4.4 & \phantom{-}0.6 & 10.7 \\
Always-foreign     & 15.3 & \phantom{1}4.8 & \phantom{-}1.1 & & 12.3 & \phantom{-}3.5 & \phantom{-}6.0 & 11.0 \\
Kept SOE           & 17.5 & 13.2 & -0.3 & & 10.4 & -0.6 & -2.2 & \phantom{1}9.9 \\
Durable privatizer & 16.8 & 15.0 & \phantom{-}0.5 & & 11.7 & \phantom{-}1.6 & -1.6 & 10.7 \\
\bottomrule
\end{tabular}
\begin{tablenotes}\footnotesize
\item \textit{Notes:} Mean firm-level log change between 1999 and 2008 divided by the number of years, restricted to firms present in every year so that composition is held fixed. $y,k,l,m$ are log deflated output, capital, labor, and materials.
\end{tablenotes}
\end{threeparttable}
\end{table}

\paragraph{How the kept SOEs raise capital and labor efficiency.}
The raw output and inputs in Table~\ref{tab:soe_within_growth} corroborate this restructuring. The kept SOEs raise output by $10.4\%$ a year and materials by $9.9\%$ while their capital stock barely moves ($-0.6\%$) and their workforce shrinks ($-2.2\%$): they add output by drawing in intermediate inputs, not by accumulating capital or labor. Always-private firms grow the opposite way, expanding output ($13.6\%$), capital ($8.0\%$), and labor ($3.0\%$) together. The kept SOEs thus raise their capital- and labor-augmenting productivity by drawing more output from a capital stock that does not grow and a workforce that shrinks, while materials, whose augmenting productivity is roughly flat, scale with output.

\section{Robustness to Capital Adjustment Frictions} \label{sec:capwedge}

My cost-share identities treat capital as a static input whose user cost equals its shadow price. Adjustment frictions, irreversibility, or financing constraints would open a wedge between the two and could contaminate the estimated factor-augmenting productivities. \citet{gottardo2026production} shows that this contamination could operate through one channel only: the output system~(\ref{eq:beforegmm}) holds however capital was accumulated, provided capital is priced at its shadow value, so a friction enters the estimation as mismeasurement of the cost shares, not as a failure of the equations. The component of this mismeasurement orthogonal to the lagged instruments is harmless. It enlarges the residual variance without biasing the estimates. The operative threat is the systematic component, a persistent wedge between user cost and shadow price, and it is not recoverable within the estimation and must be supplied from outside. \citet{gottardo2026production} reaches the same conclusion. The purging step proposed there corrects noise in the cost of capital but not a bias in its level, and input-market wedges of exactly this kind enter the cost shares only at externally supplied values. I therefore treat the wedge as a primitive, fixing its value and bounding its effect in two forms below: a uniform proportional wedge, and an ownership-specific one the capital-year effects cannot absorb. Here, I model the first as a constant proportional wedge $(1+\tau)$ on the capital first-order condition, $MC_{jt}\,\partial Q_{jt}/\partial K_{jt}=P^K_{jt}\,(1+\tau)$, while labor and materials remain flexible. The implied structural cost shares $S^{*X}_{jt}$, defined over the true cost $TC^*_{jt}=P^K_{jt}(1+\tau)K_{jt}+P^L_{jt}L_{jt}+P^M_{jt}M_{jt}$, still sum to one, so the output system~(\ref{eq:beforegmm}) holds with $S^{*X}_{jt}$ in place of the observed shares $S^X_{jt}$. Writing $d_{jt}\equiv\ln(TC^*_{jt}/TC_{jt})=\ln(1+\tau S^K_{jt})$ for the log cost gap, the wedge-augmented system, before pseudo-differencing, is
\begin{equation}
\left\{
\begin{aligned}
& y_{jt} = \nu k_{jt} + \nu \frac{\sigma}{1-\sigma} s^K_{jt} + \nu \frac{\sigma}{1-\sigma}\big(\ln(1+\tau) - d_{jt}\big) + \nu \omega^K_{jt} + \varepsilon_{jt} \\
& y_{jt} = \nu l_{jt} + \nu \frac{\sigma}{1-\sigma} s^L_{jt} - \nu \frac{\sigma}{1-\sigma} d_{jt} + \nu \omega^L_{jt} + \varepsilon_{jt} \\
& y_{jt} = \nu m_{jt} + \nu \frac{\sigma}{1-\sigma} s^M_{jt} - \nu \frac{\sigma}{1-\sigma} d_{jt} + \nu \omega^M_{jt} + \varepsilon_{jt}
\end{aligned}
\right. .
\label{eq:wedgesys}
\end{equation}
Comparing with the wedge-free system~(\ref{eq:beforegmm}), the friction enters in only two ways: the constant $\nu\frac{\sigma}{1-\sigma}\ln(1+\tau)$ in the capital equation, which the capital-year effects absorb after pseudo-differencing, and the common cost-gap term $-\nu\frac{\sigma}{1-\sigma}d_{jt}$, whose influence is limited by the small cross-firm dispersion of pseudo-differenced capital shares since $d_{jt}\approx\tau S^K_{jt}$. The full derivation is in \ref{Appendix:wedge}.\footnote{The same construction accommodates labor-market frictions. \ref{Appendix:wedge} gives the joint capital--labor generalization, of which the capital-only analysis here is the special case $\tau^L=0$.} A uniform wedge can therefore move only the level of capital-augmenting productivity, which the year effects absorb and which is not separately identified from them. The growth rates and the cross-factor and cross-ownership contrasts the analysis reports are unaffected. Table~\ref{tab:capwedge} re-estimates the system with the wedge-adjusted structural shares over $\tau\in[0,0.30]$ and confirms this numerically. A $30\%$ increase in the effective capital user cost moves the elasticity of substitution only from $0.384$ to $0.388$ and returns to scale from $0.975$ to $0.977$, while no within-firm privatization wedge moves by more than $0.002$ and the persistence parameters and factor-augmenting growth rates are nearly unchanged.\footnote{The wedge is not separately estimable from the production moments: freed as a parameter, its level is absorbed by the capital-year effects and only a weak nonlinear curvature identifies it, so its standard error spans the entire economically relevant range. I therefore fix it on a grid rather than estimate it.}

\begin{table}[htbp]
\centering
\begin{threeparttable}
\caption{Robustness to a Capital Adjustment-Cost Wedge.}
\label{tab:capwedge}
\begin{tabular}{lccccccc}
\toprule
$\tau$ & 0.00 & 0.05 & 0.10 & 0.15 & 0.20 & 0.25 & 0.30 \\
\midrule
$\sigma$ & 0.384 & 0.384 & 0.385 & 0.386 & 0.386 & 0.387 & 0.388 \\
$\nu$ & 0.975 & 0.975 & 0.975 & 0.976 & 0.976 & 0.976 & 0.977 \\
$\rho^K$ & 0.806 & 0.806 & 0.806 & 0.807 & 0.807 & 0.807 & 0.807 \\
\midrule
\multicolumn{8}{l}{\textit{SOE$\to$Private/Foreign privatization wedges $\beta^{X,w}_{SP}-\pi^X_{SP}$}} \\
$\omega^K$ early & \phantom{-}0.021 & \phantom{-}0.021 & \phantom{-}0.021 & \phantom{-}0.021 & \phantom{-}0.021 & \phantom{-}0.021 & \phantom{-}0.021 \\
$\omega^K$ late & -0.049 & -0.049 & -0.049 & -0.049 & -0.049 & -0.049 & -0.049 \\
$\omega^L$ early & \phantom{-}0.023 & \phantom{-}0.024 & \phantom{-}0.024 & \phantom{-}0.024 & \phantom{-}0.024 & \phantom{-}0.024 & \phantom{-}0.024 \\
$\omega^L$ late & -0.011 & -0.011 & -0.011 & -0.011 & -0.011 & -0.010 & -0.010 \\
$\omega^M$ early & -0.017 & -0.017 & -0.017 & -0.017 & -0.017 & -0.017 & -0.017 \\
$\omega^M$ late & -0.016 & -0.016 & -0.016 & -0.016 & -0.016 & -0.016 & -0.017 \\
\bottomrule
\end{tabular}
\begin{tablenotes}
\footnotesize
\item \textit{Notes:} At $\tau=0$ the estimates reproduce Table~\ref{tab:gmm_priv}. The elasticity of substitution, returns to scale, and the early/late privatization wedges are stable across the entire grid.
\end{tablenotes}
\end{threeparttable}
\end{table}

\paragraph{An ownership-specific wedge.} A uniform wedge is the innocuous case. The wedge that can bias the estimates is one that is persistent and correlated with the instruments, neither differenced away nor absorbed by the year effects, and the leading candidate is an ownership-specific capital cost. If state firms borrowed below market under soft budget constraints, their user cost overstates their effective cost of capital, and the distortion switches off precisely when they leave state control, an ownership-specific instance of the firm-level capital wedges that generate factor misallocation \citep{hsieh2009misallocation}. \citet{brandttombezhu2013factor} document a state-capital subsidy of this kind directly, which fixes the sign of the wedge. State capital was subsidized, not penalized. I therefore set $\tau^K_{jt}=\tau_{\mathrm{SOE}}\,D^{\mathrm{SOE}}_{jt}$ and re-estimate over the subsidy range $\tau_{\mathrm{SOE}}\in[-0.3,\,0]$, a discount of up to $30\%$ on the user cost of state capital. Table~\ref{tab:capwedge_soe} reports the grid. The technology parameters barely move at any subsidy, with $\hat\sigma$ fixed at $0.384$ and $\hat\nu$ within $0.001$ of $0.975$, and the non-state capital-augmenting advantage over kept SOEs narrows from $0.160$ to $0.118$ at a $30\%$ subsidy while staying positive throughout. The capital privatization wedges respond most, and they move in one direction: deepening the subsidy lowers the early wedge $\beta^{K,\mathrm{ea}}_{SP}-\pi^K_{SP}$, already marginal at baseline ($+0.021$), to $-0.092$ at $30\%$, and the late wedge from $-0.049$ to $-0.090$. Mechanically, the deeper the subsidy, the larger the rise in the firm's effective capital share when it leaves state control, and the less of the observed transition remains to be attributed to capital-augmenting productivity. The baseline value is thus an upper bound. Under any subsidy the early wedge lies below it, turning negative once the subsidy reaches $10\%$, and the late wedge is negative throughout. A state-capital subsidy can therefore only strengthen the finding that privatization does not raise within-firm capital-augmenting productivity. Part of the measured transition is then the repricing of formerly subsidized capital at market rates, consistent with the markup evidence of \ref{Appendix:markup} that the ownership change shows up in prices rather than in factor efficiency.

\begin{table}[htbp]
\centering
\begin{threeparttable}
\caption{Robustness to an SOE-Specific Capital Subsidy.}
\label{tab:capwedge_soe}
\begin{tabular}{lcccc}
\toprule
$\tau_{\mathrm{SOE}}$ & 0.00 & -0.10 & -0.20 & -0.30 \\
\midrule
$\sigma$ & 0.384 & 0.384 & 0.384 & 0.384 \\
$\nu$ & 0.975 & 0.975 & 0.975 & 0.974 \\
$\alpha^K_P$ & \phantom{-}0.160 & \phantom{-}0.148 & \phantom{-}0.134 & \phantom{-}0.118 \\
\midrule
$\beta^{K,\text{ea}}_{SP}-\pi^K_{SP}$ & \phantom{-}0.021 & -0.012 & -0.049 & -0.092 \\
 & (0.0119) & (0.0119) & (0.0121) & (0.0125) \\
$\beta^{K,\text{la}}_{SP}-\pi^K_{SP}$ & -0.049 & -0.061 & -0.075 & -0.090 \\
 & (0.0116) & (0.0116) & (0.0117) & (0.0118) \\
\bottomrule
\end{tabular}
\begin{tablenotes}
\footnotesize
\item \textit{Notes:} Each column re-estimates the main system with an SOE-specific effective capital user cost $(1+\tau_{\mathrm{SOE}}D^{\mathrm{SOE}}_{jt})P^K$; $\tau_{\mathrm{SOE}}<0$ subsidizes state capital, the sign documented by \citet{brandttombezhu2013factor}. At $\tau_{\mathrm{SOE}}=0$ the estimates reproduce Table~\ref{tab:gmm_priv}. Standard errors clustered by firm, of the within-firm wedges, in parentheses. The wedges fall as the subsidy deepens, so the baseline near-zero early capital effect is an upper bound.
\end{tablenotes}
\end{threeparttable}
\end{table}

\section{Conclusion} \label{sec:conclusion}

This article develops a method for estimating a CES production function with firm-level capital-, labor-, and material-augmenting productivity. Built on static cost minimization and dynamic-panel GMM, it recovers each factor-augmenting productivity, its persistence, and its dependence on ownership and privatization, without assumptions on firms' product-market conduct. I apply it to Chinese manufacturing from 1998 to 2008 during the SOE reform.

Technological change over the decade was strongly biased toward capital and labor. Capital-augmenting productivity grew fastest, at about $16.9\%$ a year, closely followed by labor at $14.2\%$, while material-augmenting productivity was flat. Ownership shapes the level of productivity, with non-state firms above SOEs on capital and labor but below them on materials. Privatization itself works through labor and, on the full panel, only at the transition. Set against the SOEs the state kept, it raises labor-augmenting productivity in the early years after the transition. The gain fades on the full panel but is large and persistent among the 1998-cohort incumbents. Across industries it is strongest in the transition window in the most overstaffed sectors, machinery and transport and chemicals, and it persists into the late window in machinery alone. An event study of the recovered productivity confirms this timing for labor (\ref{Appendix:eventstudy}). Privatization leaves capital-augmenting productivity no higher and modestly lowers material-augmenting productivity, and is instead associated with higher markups (\ref{Appendix:markup}).

The state sector's catch-up therefore runs through the restructuring of the kept SOEs and, for labor, the selective exit of the most overstaffed plants, not through ownership transfer, which reconciles the higher productivity of private firms with the small aggregate effect of privatization. Privatization is one margin of the reform, and the null documented here concerns that margin, not the reform as a whole. The framework applies directly to later periods and to other economies undergoing rapid structural transformation.
\clearpage

\singlespacing
\setlength\bibsep{0pt}

\renewcommand{\refname}{References}
\bibliography{ref}
\bibliographystyle{apalike}
\clearpage
\appendix
\setcounter{section}{0}
\renewcommand{\thesection}{Appendix \Alph{section}}
\numberwithin{table}{section}
\numberwithin{figure}{section}
\renewcommand{\thetable}{\Alph{section}.\arabic{table}}
\renewcommand{\thefigure}{\Alph{section}.\arabic{figure}}
\section{Identification of Productivity Process} \label{Appendix:B}

This appendix solves the two cross-moments of Section~\ref{sec:Estimation} for the variance of the measurement error and the variance of the productivity shock. Input-factor subscripts are suppressed throughout.

The variance of the measurement error can be identified by the following:

\begin{equation*}
\begin{aligned}
	\mathbb{E}[\widetilde{y}_{t} \ \widetilde{y}_{t-1}] & =  \mathbb{E}[(\rho \widetilde{y}_{t-1} + \nu\xi_t +\nu z_t+ \varepsilon_t - \rho \varepsilon_{t-1}) \ \widetilde{y}_{t-1}] \\
	& = \rho \mathbb{E}[\widetilde{y}_{t-1} \widetilde{y}_{t-1} ] - \rho \sigma_{\varepsilon}^2 +\nu \mathbb{E}[z_t \widetilde{y}_{t-1}]
\end{aligned}
\end{equation*}
\begin{equation*}
	\sigma_{\varepsilon}^2 = \frac{\rho \mathbb{E}[\widetilde{y}_{t-1} \widetilde{y}_{t-1} ] - \mathbb{E}[\widetilde{y}_{t} \ \widetilde{y}_{t-1}] +\nu \mathbb{E}[z_t \widetilde{y}_{t-1}]}{\rho}
\end{equation*}
The variance of the productivity shock can be identified by the following:

\begin{equation*}
\begin{aligned}
	\mathbb{E}[\widetilde{y}_{t} \ \widetilde{y}_{t}] & =  \mathbb{E}[(\rho \widetilde{y}_{t-1} + \nu\xi_t +\nu z_t+ \varepsilon_t - \rho \varepsilon_{t-1}) \ (\rho \widetilde{y}_{t-1} + \nu\xi_t +\nu z_t+ \varepsilon_t - \rho \varepsilon_{t-1})] \\
	& = \rho^2 \mathbb{E}[\widetilde{y}_{t-1} \widetilde{y}_{t-1} ] + (1-\rho^2)\sigma^2_{\varepsilon} + \nu^2 \sigma^2_\xi + 2\nu\rho \mathbb{E}[z_t \widetilde{y}_{t-1}] + \nu^2 \mathbb{E}[z_t^2]
\end{aligned}
\end{equation*}
\begin{equation*}
	\sigma^2_\xi = \frac{\mathbb{E}[\widetilde{y}_{t} \widetilde{y}_{t} ] - \rho^2 \mathbb{E}[\widetilde{y}_{t-1} \widetilde{y}_{t-1} ] - (1-\rho^2) \sigma_{\varepsilon}^2 - 2\nu\rho \mathbb{E}[z_t \widetilde{y}_{t-1}] - \nu^2 \mathbb{E}[z_t^2]}{\nu^2}
\end{equation*}

\section{GLS-based Recovery of Factor-Augmenting Productivity} \label{Appendix:gls}
This appendix derives the GLS-based filter of Section~\ref{sec:Estimation}, which recovers firm-year factor-augmenting productivity from the residual series $\widetilde{y}_{Xjt} = \nu\omega^X_{jt} + \varepsilon_{jt}$ by treating it as a state-space model.

Here I derive the estimation process of $\omega^X_{jt}$ for a single input factor, suppressing the input-factor superscript and the firm subscripts for brevity. From Equation (\ref{eq:ar1}), I write down the recursive form of $\omega_t$ and solve for $\omega_1$, $\omega_2$, $\dots$, $\omega_{t-1}$ as functions of $\omega_t$ and $\xi_t$, $\xi_{t-1}$, $\xi_{t-2}$, $\dots$:

\begin{equation}
\begin{bmatrix}
\omega_1 \\
\omega_2 \\
\vdots \\
\omega_{t-1} \\
\omega_t
\end{bmatrix}
=
\begin{bmatrix}
\rho^{-t+1} \omega_t \\
\rho^{-t+2} \omega_t \\
\vdots \\
\rho^{-1} \omega_t \\
\omega_t
\end{bmatrix}
-
\begin{bmatrix}
\rho^{-1} (\xi_2+z_2) + \rho^{-2} (\xi_3+z_3) + \cdots + \rho^{-t+1} (\xi_t + z_t) \\
\rho^{-1} (\xi_3 +z_3) + \rho^{-2} (\xi_4 + z_4) + \cdots + \rho^{-t+2} (\xi_t + z_t)\\
\vdots \\
\rho^{-1} (\xi_t+z_t) \\
0
\end{bmatrix}.
\label{eq:omega_mat}
\end{equation}

Combining Equation (\ref{eq:omega_mat}) with $\widetilde{y}_{t} = \nu\omega_t + \varepsilon_t$, I have

\begin{equation}
\begin{bmatrix}
\widetilde{y}_1 \\
\widetilde{y}_2 \\
\vdots \\
\widetilde{y}_{t-1} \\
\widetilde{y}_t
\end{bmatrix}
=
\nu \begin{bmatrix}
 \rho^{-t+1} \\
 \rho^{-t+2} \\
\vdots \\
 \rho^{-1} \\
1
\end{bmatrix}
\omega_t
-
\nu \begin{bmatrix}
 \rho^{-1} &  \rho^{-2} & \cdots &  \rho^{-t+1} \\
0 &  \rho^{-1} & \cdots &  \rho^{-t+2} \\
\vdots & \vdots & \ddots & \vdots \\
0 & 0 & \cdots &  \rho^{-1} \\
0 & 0 & \cdots & 0
\end{bmatrix}
\begin{bmatrix}
\xi_2 + z_2\\
\xi_3 + z_3 \\
\vdots \\
\xi_{t-1} + z_{t-1} \\
\xi_t + z_t
\end{bmatrix}
+
\begin{bmatrix}
\varepsilon_1 \\
\varepsilon_2 \\
\vdots \\
\varepsilon_{t-1} \\
\varepsilon_t
\end{bmatrix}.
\label{eq:tilde_y_t}
\end{equation}
Since the deterministic shifters $z_2,\dots,z_t$ (each combining the year fixed effect and the ownership coefficients) are identified in the GMM step, I subtract the terms containing them from both sides, yielding a matrix form of $\mathbf{\widetilde{\widetilde{y}}_t}$:
\begin{equation}
    \mathbf{\widetilde{\widetilde{y}}_t} = \mathbf{\widetilde{y}_t} + \mathbf{A_t} \mathbf{z_t}.
\label{eq:tilde_y_t_2}
\end{equation}
where $\mathbf{z_t} \equiv (z_2,\dots,z_t)'$ and $\boldsymbol{\xi}_t \equiv (\xi_2,\dots,\xi_t)'$ stack the shifters and innovations of periods $2$ through $t$, and $\mathbf{A_t}$ is the $t \times (t-1)$ matrix \begin{equation}
    \mathbf{A_t} =
\nu \begin{bmatrix}
\rho^{-1} & \rho^{-2} & \cdots &  \rho^{-t+1} \\
0 &  \rho^{-1} & \cdots &  \rho^{-t+2} \\
\vdots & \vdots & \ddots & \vdots \\
0 & 0 & \cdots &  \rho^{-1} \\
0 & 0 & \cdots &  0
\end{bmatrix},
\end{equation}
so that $\mathbf{A_t}\mathbf{z_t}$ is $t \times 1$, conformable with $\mathbf{\widetilde{y}_t}=(\widetilde{y}_1,\dots,\widetilde{y}_t)'$.

I combine Equation (\ref{eq:tilde_y_t}) and Equation (\ref{eq:tilde_y_t_2}) as a matrix form:

\begin{equation}
    \mathbf{\widetilde{\widetilde{y}}_t} = \mathbf{X}_t^* \omega_t + \mathbf{e}_t,
\end{equation}
where the $\mathbf{X}_t^*$ is defined as the scalar $\nu$ multiplied by the vector of inverse powers of $\rho$, and the error term is given by $\mathbf{e}_t = -\mathbf{A}_t \boldsymbol{\xi}_t + \boldsymbol{\varepsilon}_t$. The corresponding weighting matrix $\mathbf{\Omega_t}$ of GLS estimation is defined as:

\begin{equation}
   \mathbb{E}(\mathbf{e}_t \mathbf{e}_t') = \mathbf{A_t} (\mathbf{I_{t-1}} \otimes \sigma^2_{\xi}) \mathbf{A_t}^{'} + \sigma^2_{\varepsilon} \mathbf{I_t} = \mathbf{\Omega_t}.
\end{equation}

Then I can get each one of the three factor-augmenting productivities by estimating the following equation\footnote{The productivity of the first year cannot be estimated in many econometric models due to a lack of prior information and the nature of dynamic processes as first-order Markov process. Therefore, the productivity of firms in the year 1998 will not be reported in the subsequent results.}:

\begin{equation}
    \omega_{t|t} = \left( \mathbf{X_t}^{*'} \mathbf{\Omega_t}^{-1} \mathbf{X_t}^{*} \right)^{-1} \mathbf{X}_t^{*'} \mathbf{\Omega_t}^{-1} \mathbf{\widetilde{\widetilde{y}}_t}, \quad \text{for} \quad t = 2, 3, \dots, T.
\end{equation}

\section{Year Fixed Effects} \label{Appendix:yearfe}
Table~\ref{tab:time_dummies} reports the year fixed effects $\gamma^X_t$, which capture economy-wide shifts in factor-augmenting productivity that are common across firms after controlling for ownership selection. From $1999$ onward, all three factor-specific time effects trend upward. $\gamma^K_t$ and $\gamma^L_t$ dip in $2001$ and $2004$ but grow rapidly through the mid-$2000$s, while $\gamma^M_t$ increases more modestly. The pattern in $2008$ reflects the onset of the global financial crisis: $\gamma^K_t$ and $\gamma^L_t$ both pull back while $\gamma^M_t$ continues to rise. The clustered standard errors are small relative to the estimates, and the time effects track SOE reforms, WTO-led export growth, and the global downturn.

\begin{table}[!ht]
\centering
\begin{threeparttable}
\caption{GMM Estimates: Year Fixed Effects $\gamma^X_t$.}
\label{tab:time_dummies}
\begin{tabular}{ccccccc}
\toprule
Year & \multicolumn{2}{c}{$\gamma_t^K$} & \multicolumn{2}{c}{$\gamma_t^L$} & \multicolumn{2}{c}{$\gamma_t^M$} \\
\cmidrule(lr){2-3} \cmidrule(lr){4-5} \cmidrule(lr){6-7}
 & Est. & SE & Est. & SE & Est. & SE \\
\midrule
  1999 & 0.591 & (0.019) & 1.063 & (0.017) & 0.330 & (0.010) \\
  2000 & 0.638 & (0.020) & 1.081 & (0.017) & 0.335 & (0.010) \\
  2001 & 0.615 & (0.019) & 1.061 & (0.016) & 0.320 & (0.010) \\
  2002 & 0.707 & (0.020) & 1.143 & (0.017) & 0.335 & (0.010) \\
  2003 & 0.765 & (0.021) & 1.180 & (0.017) & 0.342 & (0.010) \\
  2004 & 0.739 & (0.021) & 1.157 & (0.017) & 0.348 & (0.010) \\
  2005 & 0.801 & (0.022) & 1.227 & (0.017) & 0.356 & (0.010) \\
  2006 & 0.838 & (0.022) & 1.262 & (0.018) & 0.335 & (0.010) \\
  2007 & 0.861 & (0.022) & 1.285 & (0.017) & 0.358 & (0.010) \\
  2008 & 0.755 & (0.022) & 1.204 & (0.016) & 0.384 & (0.010) \\
\bottomrule
\end{tabular}
\begin{tablenotes}
\footnotesize
\item \textit{Notes:} Year fixed effects $\gamma^X_t$ ($1998$ omitted) from the main specification (kept-SOE baseline, ownership dated $t$), estimated on $N=1{,}262{,}501$ firm-years. Standard errors clustered by firm, in parentheses.
\end{tablenotes}
\end{threeparttable}
\end{table}

\section{Capital Adjustment-Cost Wedge: Derivation} \label{Appendix:wedge}
This appendix derives the wedge-augmented system of Section~\ref{sec:capwedge} and the identification of $\tau$. Under the wedged capital first-order condition of that section, the relevant cost concept is the true cost $TC^*_{jt}=P^K_{jt}(1+\tau)K_{jt}+P^L_{jt}L_{jt}+P^M_{jt}M_{jt}$, and repeating the cost-share manipulation of Section~\ref{sec:model} with the structural shares $S^{*X}_{jt}$ gives $y_{jt}=\nu x_{jt}+\nu\frac{\sigma}{1-\sigma}\ln S^{*X}_{jt}+\nu\omega^X_{jt}+\varepsilon_{jt}$ for each input. In the data I observe only the apparent shares $S^X_{jt}=P^X_{jt}X_{jt}/TC_{jt}$, and with the log cost gap $d_{jt}\equiv\ln(TC^*_{jt}/TC_{jt})=\ln(1+\tau\,S^K_{jt})\approx\tau\,S^K_{jt}$ the structural log shares $s^{*X}_{jt}\equiv\ln S^{*X}_{jt}$ relate to the observed ones by $s^{*K}_{jt}=s^K_{jt}+\ln(1+\tau)-d_{jt}$, $s^{*L}_{jt}=s^{L}_{jt}-d_{jt}$ and $s^{*M}_{jt}=s^{M}_{jt}-d_{jt}$. Substituting them into the three output equations returns system~(\ref{eq:wedgesys}).

Because the structural shares are observable once $\tau$ is fixed, I estimate the system with $s^{*X}_{jt}(\tau)$ in place of the observed $s^X_{jt}$, which keeps the composite residual clean. Pseudo-differencing $(1-\rho^X L)$ to remove the AR(1) productivity gives, for each factor,
\begin{equation*}
y_{jt}=\rho^X y_{j,t-1}+\nu(x_{jt}-\rho^X x_{j,t-1})+\nu\tfrac{\sigma}{1-\sigma}\big(s^{*X}_{jt}-\rho^X s^{*X}_{j,t-1}\big)+\nu z^X_{jt}+u^X_{jt},\qquad u^X_{jt}=\nu\xi^X_{jt}+(\varepsilon_{jt}-\rho^X\varepsilon_{j,t-1}).
\end{equation*}
For a constant wedge, $s^{*K}_{jt}-\rho^K s^{*K}_{j,t-1}=(s^K_{jt}-\rho^K s^K_{j,t-1})+\ln(1+\tau)(1-\rho^K)-(d_{jt}-\rho^K d_{j,t-1})$. The constant $\nu\frac{\sigma}{1-\sigma}\ln(1+\tau)(1-\rho^K)$ is collinear with the capital-specific year effect $\gamma^K_t$ and is absorbed by it, so $\sigma$, $\nu$, and $\rho^X$ do not depend on the level of the wedge. The only remaining effect is the firm-varying curvature $-(d_{jt}-\rho^K d_{j,t-1})$ with $d_{jt}\approx\tau S^K_{jt}$, which enters linearly in $\tau$ with a coefficient governed by the small cross-firm dispersion of pseudo-differenced capital shares. The estimates therefore move only slightly as $\tau$ varies. To estimate $\tau$, I restore the moment $\mathbb{E}[(s^K_{j,t-1})'u^K_{jt}]=0$ that the just-identified baseline leaves out. The clean residual makes the lagged share a valid instrument, and $\tau$ is identified because the regressor $s^{*K}_{jt}(\tau)$ depends on it through $d_{jt}\approx\tau S^K_{jt}$. As that dependence is weak, the wedge is poorly identified.

\paragraph{Joint capital--labor wedges.} The construction extends directly to a labor-market friction. With proportional wedges $(1+\tau^K)$ and $(1+\tau^L)$ on capital and labor, the true cost is $TC^*_{jt}=P^K_{jt}(1+\tau^K)K_{jt}+W_{jt}(1+\tau^L)L_{jt}+P^M_{jt}M_{jt}$, the cost gap becomes $d_{jt}=\ln(1+\tau^K S^K_{jt}+\tau^L S^L_{jt})$, and the structural log shares are $s^{*K}_{jt}=s^K_{jt}+\ln(1+\tau^K)-d_{jt}$, $s^{*L}_{jt}=s^L_{jt}+\ln(1+\tau^L)-d_{jt}$, and $s^{*M}_{jt}=s^M_{jt}-d_{jt}$. The estimating system is otherwise unchanged: each constant $\ln(1+\tau^X)$ is absorbed by the corresponding factor's year effect, so $\sigma$, $\nu$, and the persistence parameters remain invariant to the levels of the wedges, while a labor wedge is identified, alongside the capital wedge, by adding a labor-side moment not already used in the just-identified baseline, such as the twice-lagged share $\mathbb{E}[(s^L_{j,t-2})'u^L_{jt}]=0$. The capital-only specification of the main text is the special case $\tau^L=0$. The wedges can be made ownership-specific, e.g.\ $\tau^X_{jt}=\tau^X_0+\tau^X_{\mathrm{SOE}}D^{\mathrm{SOE}}_{jt}$. The framework then speaks to SOE-specific capital- and labor-market distortions.

\bigskip

\section{Data Construction}\label{Appendix:data}

This article follows the dataset treatment and construction of \citet{jaumandreu2018cost}.

\subsection*{Data treatment}

In the raw data, the same firm can appear at different times with different identifiers. Linking these observations is essential to correctly track each firm's time sequence and identify shutdowns. After linking, I observe reasonable rates of economic entry and exit, along with survey additions. Many additions are likely due to firms growing large enough to qualify for the survey, but some also include firms with sales exceeding 5 million RMB. Moreover, the ratio of data aggregates to industry GDP in the \textit{China Statistical Yearbook} shows an increasing trend. Statistical coverage likely improved over the period.

I also ensure data consistency by dropping abnormal observations and using each firm's longest sequence of complete data lasting more than one year. The cleaned dataset retains 84\% of the firms and 74\% of the original observations.

\subsection*{Variables definition and construction}

\paragraph{Revenue} Revenue after taxes, at current prices, as reported by the firm.

\paragraph{Price of output} Output price index of the two-digit industry the firm belongs to, taken from China Statistical Yearbook (CSY).

\paragraph{Capital} Real capital stock is constructed as follows. Firms report the value of their capital stock at original purchase prices and at original purchase prices less accumulated depreciation. Using these nominal values, I estimate a sequence of real investments and real capital stock for the starting year. The capital stock is then calculated using the perpetual inventory method, assuming a yearly depreciation rate of 9\%.

For firms established before 1998, I use a method similar to that of \citet{brandt2012creative}. Specifically, I estimate a yearly nominal rate of investment in fixed assets at the two-digit industry level based on data from 1998--2003. I assume that capital accumulates at this constant rate from the firm's inception. Using this, I estimate the initial capital stock, deflate it, and compute the real stock for the first year. The investment deflator is derived from \citet{brandt2008industrial} and updated using the Fixed Asset Investment price index from the \textit{China Statistical Yearbook}.

\paragraph{User cost of capital}

Computed following \citet{doraszelski2018measuring} as $P_{It}(r_{t} + \delta - \text{CPI}_t)$, where $P_{It}$ is a province-by-year investment price index, $r_{t}$ is the nominal interest rate, $\delta$ is the depreciation rate, and $\text{CPI}_t$ is the rate of inflation as measured by the consumer price index. The same $9\%$ depreciation rate enters both the perpetual-inventory capital stock and the flow user cost. The constructed user cost is common to all firms within a province and year, so cross-firm differences in the capital cost share come from their capital stocks rather than from firm-specific costs of funds. A wedge between this constructed user cost and a firm's effective cost of capital is treated directly in Section~\ref{sec:capwedge}, as a uniform proportional wedge and as an ownership-specific state-capital subsidy, with its effect on the technology parameters and the privatization effect bounded there.

\paragraph{Cost of materials} The estimation of intermediate consumption in production is carried out as follows. The survey's definition of intermediate inputs includes direct materials, intermediate inputs used in production, inputs for management, inputs for business operations (sales costs), and financial expenses. However, as my focus is on variable costs, the inclusion of general management expenses, sales costs, and financial costs is not ideal.

Instead, I use manufacturing costs, which comprise materials, labor costs, and depreciation of capital during the production process. From these manufacturing costs, I impute the wage bill and the depreciation of capital. For the years 2004 to 2007, I utilize detailed information on the structure of intermediate inputs to perform this calculation. For other years, I assume the same proportions as in 2004--2007. Concretely, materials cost equals manufacturing cost net of the imputed wage bill and depreciation of capital, with the wage-bill and depreciation shares of manufacturing cost taken from the detailed 2004--2007 input-structure module and held fixed at their firm-level values for the remaining years.

\paragraph{Price of materials} To estimate a price index for the intermediate consumption of the industry to which a firm belongs, I follow the approach of \citet{brandt2012creative}. Specifically, I compute a weighted average of the output prices for the industries supplying inputs to the considered industry. The weights are derived from the 2002 Input-Output table, which includes 42 sectors.

The two-digit manufacturing price indices are sourced from the China Statistical Yearbook (CSY). For agriculture, construction, transportation, retail, wholesale, and certain service sectors, prices are calculated by comparing GDP at current prices and constant prices, as documented in the \textit{Collection of Statistical Material from 1949 to 2009}.

\paragraph{Materials} Cost of materials divided by the price of materials.

\paragraph{Wage bill} I add up wages, unemployment insurance premium, pension and medical insurance premium, housing mutual fund and total welfare fees. It should be taken into account that firms only began to report retirement and health insurance in 2003, and housing benefits in 2004. Any common level shift from this change in benefit reporting is absorbed by the factor-specific year effects $\gamma^L_t$ in the productivity process, so it does not bias the estimated persistence or privatization coefficients.

\paragraph{Employment} Total number of employees, which includes all the full-time production and nonproduction workers, as reported by the firm. It excludes part-time and casual workers.

\paragraph{Wage} Wage bill divided by employment.

\paragraph{Ownership classification} Firms are classified into four ownership types (state-owned, private, collective, and foreign), based on the identity of the majority shareholder, using a 50\% equity threshold to determine control, following \citet{brandt2012creative} and \citet{hsieh2015grasp}, as described in Section~\ref{sec:priv_data}. Each of the four types is carried separately throughout the analysis. Foreign firms are pooled with private firms only in the destination classification of the durable privatizers ($\mathrm{POST}^{SP}$).

\begin{table}[ht]
\centering
\renewcommand{\arraystretch}{1.2}
\setlength{\tabcolsep}{8pt}
\begin{threeparttable}
\caption{Summary Statistics of Production Variables and Price Deflators.}
\label{tab:summary_statistics}
\begin{tabular}{lcc}
\toprule
Variable & Mean & Standard Deviation \\
\midrule
Revenue  & 83.393 & 721.860 \\
Output Price  & 1.0135 & 0.1790 \\
Capital & 24.073 & 307.880 \\
Price of Investment  & 1.2447 & 0.1823 \\
Employment  & 238.08 & 806.33 \\
Wage Bill  & 4.554 & 32.309 \\
Materials  & 66.150 & 610.030 \\
Price of Materials  & 1.1264 & 0.2013 \\
\bottomrule
\end{tabular}
\begin{tablenotes}
\footnotesize
\item \textit{Notes:} Firm-level means and standard deviations over the $1{,}659{,}612$ firm-year analysis panel, $1998$--$2008$. Revenue, capital, wage bill, and materials are in millions of RMB, deflated by the corresponding industry price indices; output, investment, and material prices are the deflators, normalized to the sample base. Employment is the firm headcount.
\end{tablenotes}
\end{threeparttable}
\end{table}

\subsection*{Industry correspondence}

I group the two-digit industry codes into ten sectors largely following \citet{jaumandreu2018cost}, with two minor differences dictated by identification: paper (code 22) and printing (code 23) are kept as separate sectors rather than pooled, and transport equipment (code 37) is grouped with machinery (codes 35--36). The full correspondence is reported below.

\begin{table}[htbp]
\centering
\begin{threeparttable}
\caption{Industry Classification and Corresponding Two-Digit Codes.}
\label{tab:industry_classification}
\begin{tabular}{ll}
\toprule
Sector & Constituent Two-Digit CIC Codes \\ \midrule
1. Food, drink and tobacco & 13. Agricultural and by-product proc. \\
                           & 14. Food manufacturing \\
                           & 15. Beverage manufacturing \\
                           & 16. Tobacco products \\ \addlinespace
2. Textile, leather and shoes & 17. Textile \\
                              & 18. Apparel, shoes, and hat manuf. \\
                              & 19. Leather, fur, and coat prod. manuf. \\ \addlinespace
3. Timber and furniture & 20. Wood proc., and other wood prod. \\
                        & 21. Furniture manufacturing \\ \addlinespace
4. Paper & 22. Paper making and paper products \\ \addlinespace
5. Printing & 23. Printing and recording media reprod. \\ \addlinespace
6. Chemical products & 26. Chemical materials and products \\
                      & 27. Pharmaceutical \\
                      & 28. Chemical fiber \\
                      & 29. Rubber products \\
                      & 30. Plastic products \\ \addlinespace
7. Non-metallic minerals & 31. Non-metallic minerals products \\ \addlinespace
8. Metals and metal products & 32. Ferr. metal smelting and rolling proc. \\
                             & 33. Non-ferrous metal rolling processing \\
                             & 34. Metal products \\ \addlinespace
9. Machinery and transport equipment & 35. General machinery manufacturing \\
                                     & 36. Special machinery manufacturing \\
                                     & 37. Transportation equipment manuf. \\ \addlinespace
10. Electronics & 39. Electrical machinery and equipment \\
                & 40. Elec. commun. equip. and computer \\
                & 41. Instruments, meters, and office machinery \\ \bottomrule
\end{tabular}
\begin{tablenotes}\footnotesize
\item \textit{Notes:} Two-digit codes follow the Chinese Industry Classification. The ten sectors cover every firm-year in the analysis panel; two-digit codes 24, 25, 42, and 43 do not appear in the cleaned data.
\end{tablenotes}
\end{threeparttable}
\end{table}

\section{Pooled Single-Coefficient Privatization Effect}\label{Appendix:pooled}
For comparison with the timing-split main specification (Table~\ref{tab:gmm_priv}), Table~\ref{tab:gmm_priv_pooled} reports the pooled version with a single post-transition coefficient per destination, not split into early and late windows. The pooled labor effect $\beta^L_{SP}-\pi^L=+0.008$ ($t=1.2$) is the average of the significant early gain and the null late window of the main specification. A single coefficient therefore reads as a full-panel null. This specification also carries the formal test of the private-type comparison in Section~\ref{sec:soe_catchup}, that privatized firms do not attain the never-state private terms: $\beta^X_{SP}=\alpha^X_P$ is rejected for every factor, with $\alpha^K_P-\beta^K_{SP}=0.155$ ($t=21.9$), $\alpha^L_P-\beta^L_{SP}=0.074$ ($t=15.6$), and $\alpha^M_P-\beta^M_{SP}=-0.064$ ($t=-23.0$), standard errors from the joint parameter covariance.

\begin{table}[htbp]
\centering
\begin{threeparttable}
\caption{Pooled Single-Coefficient Privatization Effect (single POST, appendix comparison).}
\label{tab:gmm_priv_pooled}
\begin{tabular}{l c c c}
\toprule
 & $\omega^{K}$ & $\omega^{L}$ & $\omega^{M}$ \\
\midrule
\multicolumn{4}{l}{\textit{Pre-privatization selection} ($\pi$)} \\
\addlinespace
Pre-privatization \ ($\pi$) & $\phantom{-}0.014$ & $\phantom{-}0.021^{**}$ & $-0.004$ \\
  & $(0.014)$ & $(0.010)$ & $(0.006)$ \\
\addlinespace
\midrule
\multicolumn{4}{l}{\textit{Post-transition terms, single POST} ($\beta_{Sd}$)} \\
\addlinespace
SOE$\to$Private/Foreign \ ($\beta_{SP}$) & $\phantom{-}0.004$ & $\phantom{-}0.029^{***}$ & $-0.017^{***}$ \\
  & $(0.013)$ & $(0.009)$ & $(0.006)$ \\
SOE$\to$Collective \ ($\beta_{SC}$) & $\phantom{-}0.028$ & $-0.036^{**}$ & $-0.036^{***}$ \\
  & $(0.023)$ & $(0.016)$ & $(0.008)$ \\
\addlinespace
\midrule
\multicolumn{4}{l}{\textit{Baseline-invariant effect} ($\beta_{Sd}-\pi$)} \\
\addlinespace
SOE$\to$Private/Foreign & $-0.010$ & $\phantom{-}0.008$ & $-0.014^{***}$ \\
  & $(0.010)$ & $(0.007)$ & $(0.004)$ \\
SOE$\to$Collective & $\phantom{-}0.014$ & $-0.057^{***}$ & $-0.032^{***}$ \\
  & $(0.021)$ & $(0.015)$ & $(0.007)$ \\
\bottomrule
\end{tabular}
\begin{tablenotes}
\footnotesize
\item \textit{Notes:} The pooled single-POST specification (one post-transition coefficient per destination, not split by timing), retained for comparison with the timing-split main Table~\ref{tab:gmm_priv}. The pooled labor effect $\beta^L_{SP}-\pi^L=+0.008$ ($t=1.2$) is the average of the significant early gain and the null late period of Table~\ref{tab:gmm_priv}. Standard errors clustered by firm. ${}^{*}\,p<0.10$, ${}^{**}\,p<0.05$, ${}^{***}\,p<0.01$.
\end{tablenotes}
\end{threeparttable}
\end{table}

\section{Event-Study Evidence on the Timing of the Privatization Effect}\label{Appendix:eventstudy}

While the structural early/late estimates of Table~\ref{tab:gmm_priv} are the paper's primary evidence on the timing of the privatization effect (Section~\ref{sec:priv_effect}), I corroborate them with an event study around the transition. The event study traces the full path of the recovered $\hat\omega^X_{jt}$. The treated group is the clean SOE$\to$private/foreign privatizers and the control is the kept SOEs (the omitted baseline and parallel-trends counterfactual of Section~\ref{sec:priv_effect}). Against event time $s=t-\tau_j$, relative to $s=-1$, I report two-way fixed effects with firm and year effects and firm-clustered leads that test for pre-trends, and the heterogeneity-robust estimator of \citet{callaway2021difference} against the never-treated kept SOEs, which avoids the forbidden comparisons that bias two-way fixed effects under staggered timing.

\begin{figure}[htbp]
\centering
\includegraphics[width=0.98\textwidth]{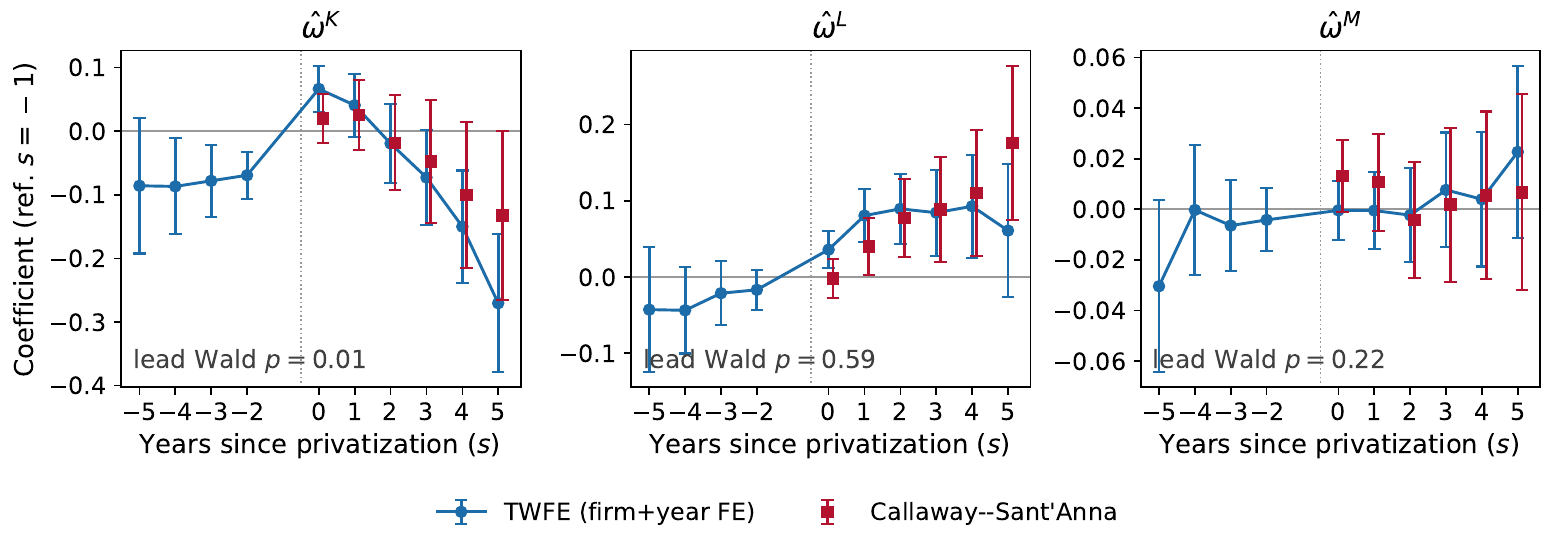}
\caption{Event study of factor-augmenting productivity around durable privatization (SOE$\to$private/foreign vs.\ kept SOEs, recovered $\hat\omega^X$). Blue circles: two-way fixed-effects coefficients with firm and year effects, $95\%$ firm-clustered intervals. Red squares: \citet{callaway2021difference} against the never-treated kept SOEs. Reference period $s=-1$. The annotation reports the joint Wald test on the pre-transition leads.}
\label{fig:event_study}
\end{figure}

Figure~\ref{fig:event_study} shows the result. Labor-augmenting productivity is flat before the transition (joint lead test $p=0.59$), rises over the first two to three years to a level gain that persists. The path reproduces the positive early and null late wedges of Panel~D. Capital-augmenting productivity rises briefly and then falls below the kept SOEs, in line with the marginal early gain and significant late decline in the structural estimates, though a pre-trend in its leads ($p=0.01$) makes the capital path only suggestive. Material-augmenting productivity is flat throughout. Cumulating the structural early/late flows through the estimated persistence traces these capital and labor paths over the first post-transition years, whereas a single pooled coefficient misses both the early labor rise and the capital reversal, so the early labor gain is not an artifact of the structural timing windows.

\section{Baseline Robustness of the Privatization Wedge}\label{Appendix:baseline}
The kept-SOE baseline conditions on survival to 2008, so a natural concern is that the privatization wedge reflects the survivor selection of the reference group. Table~\ref{tab:baseline_invariance} re-estimates the full system with the survival conditioning removed: the $\mathrm{ExitSOE}$ control is dropped and the exiting SOEs are folded into the omitted baseline, which then comprises all always-SOE firm-years regardless of survival. Both columns use the identical estimation sample. Only the normalization changes. Every wedge $\beta^{X,w}_d-\pi^X_d$ moves by less than $0.004$, well within one standard error, and no coefficient changes significance. The level structure shifts as expected against the lower pooled reference: $\alpha^K_P$ rises from $0.160$ to $0.317$ and $\hat\sigma$ moves from $0.384$ to $0.392$.

The exiting SOEs bear little resemblance to the kept ones. Their measured productivity declines toward exit, and they outnumber the kept SOEs roughly five to one in firm-years. The invariance reflects what supplies the counterfactual in the structural system. A privatizer's counterfactual dynamics come from its own lagged productivity through $\rho^X$ and from the year effects $\gamma^X_t$, which are estimated from moments over the full panel, where the exiting SOEs carry about four percent of the firm-year weight. The omitted group determines only the level against which the group coefficients are normalized, and a level shift common to $\beta^{X,w}_d$ and $\pi^X_d$ cancels in the wedge. The residual channel, that the exiters' declining calendar weight tilts $\gamma^X_t$ differentially across the privatizers' earlier pre years and later post years, is what the table bounds at $0.004$. An event study that used the SOE group alone as its control would weight the exiters five to one and inherit their declining paths as spurious pre-trends, which is why the event study of \ref{Appendix:eventstudy} benchmarks against the kept SOEs.

\begin{table}[htbp]
\centering
\begin{threeparttable}
\caption{Privatization Wedges under Alternative SOE Baselines.}
\label{tab:baseline_invariance}
{\small
\begin{tabular}{l l c c c}
\toprule
 & & Kept-SOE & All-always-SOE & Difference \\
 & & baseline (main) & baseline & \\
\midrule
\multicolumn{5}{l}{\textit{SOE$\to$Private/Foreign} ($\beta^{X,w}_{SP}-\pi^X_{SP}$)} \\
\addlinespace
$\omega^{K}$ & early & $\phantom{-}0.021^{*}$ & $\phantom{-}0.020^{*}$ & $-0.001$ \\
 & & $(0.012)$ & $(0.012)$ & \\
$\omega^{K}$ & late & $-0.049^{***}$ & $-0.052^{***}$ & $-0.003$ \\
 & & $(0.012)$ & $(0.012)$ & \\
$\omega^{L}$ & early & $\phantom{-}0.023^{***}$ & $\phantom{-}0.023^{***}$ & $-0.000$ \\
 & & $(0.008)$ & $(0.008)$ & \\
$\omega^{L}$ & late & $-0.011$ & $-0.013$ & $-0.002$ \\
 & & $(0.008)$ & $(0.009)$ & \\
$\omega^{M}$ & early & $-0.017^{***}$ & $-0.017^{***}$ & $+0.000$ \\
 & & $(0.004)$ & $(0.004)$ & \\
$\omega^{M}$ & late & $-0.016^{***}$ & $-0.016^{***}$ & $+0.000$ \\
 & & $(0.005)$ & $(0.005)$ & \\
\addlinespace
\midrule
\multicolumn{5}{l}{\textit{SOE$\to$Collective} ($\beta^{X,w}_{SC}-\pi^X_{SC}$)} \\
\addlinespace
$\omega^{K}$ & early & $-0.031$ & $-0.031$ & $-0.001$ \\
 & & $(0.039)$ & $(0.039)$ & \\
$\omega^{K}$ & late & $-0.007$ & $-0.010$ & $-0.003$ \\
 & & $(0.045)$ & $(0.045)$ & \\
$\omega^{L}$ & early & $-0.071^{**}$ & $-0.072^{**}$ & $-0.001$ \\
 & & $(0.028)$ & $(0.028)$ & \\
$\omega^{L}$ & late & $-0.073^{**}$ & $-0.076^{**}$ & $-0.003$ \\
 & & $(0.032)$ & $(0.032)$ & \\
$\omega^{M}$ & early & $\phantom{-}0.012$ & $\phantom{-}0.013$ & $+0.000$ \\
 & & $(0.013)$ & $(0.013)$ & \\
$\omega^{M}$ & late & $-0.003$ & $-0.002$ & $+0.001$ \\
 & & $(0.014)$ & $(0.014)$ & \\
\addlinespace
\midrule
\multicolumn{5}{l}{\textit{Memo}} \\
\addlinespace
$\hat\sigma$ & & $\phantom{-}0.384$ & $\phantom{-}0.392$ & $+0.008$ \\
$\hat\nu$ & & $\phantom{-}0.975$ & $\phantom{-}0.974$ & $-0.001$ \\
$\alpha^K_P$ & & $\phantom{-}0.160$ & $\phantom{-}0.317$ & $+0.157$ \\
\bottomrule
\end{tabular}}
\begin{tablenotes}
\footnotesize
\item \textit{Notes:} Both columns estimate the full system of Section~\ref{sec:Estimation} on the identical sample of $1{,}262{,}501$ firm-years; only the omitted SOE reference group changes. The main column is the kept-SOE baseline (always-SOE firms surviving to 2008, with the exiting SOEs carrying their own control $\alpha^X_E$); the comparison column drops the $\mathrm{ExitSOE}$ control and folds the exiting SOEs into the omitted baseline, so the reference group is no longer conditioned on survival. Wedge standard errors use the full parameter covariance, clustered by firm. ${}^{*}\,p<0.10$, ${}^{**}\,p<0.05$, ${}^{***}\,p<0.01$.
\end{tablenotes}
\end{threeparttable}
\end{table}

\section{Markups and Ownership}\label{Appendix:markup}

If the ownership change leaves a firm's measured factor efficiency close to the kept-SOE trajectory (Section~\ref{sec:discussion}), a
natural question is whether privatized firms differ on other margins. Following
\citet{deloecker2012markups}, who relate estimated markups to a firm's export status in a regression, I relate the markup to ownership status in the same way. I measure the markup with the CES
technology. Under cost
minimization the output elasticity of a flexible input equals returns to scale times its cost
share, $\theta^M_{jt}=\nu\,S^M_{jt}$, so the production-function markup
$\mu_{jt}=\theta^M_{jt}/(\text{materials revenue share})$ of \citet{deloecker2012markups} reduces to
$\mu_{jt}=\nu\,R_{jt}/TC_{jt}$: returns to scale times the revenue-to-cost ratio, so a markup is
recovered for every firm-year.\footnote{The estimated $\hat\omega$ and this markup are both
revenue-based objects, so I do not read the contrast between them as a split between physical
productivity and profitability. Because $\nu$ is estimated from revenue its level is not
immune to markup dispersion \citep{bond2021some}, but this bias shifts $\ln\mu_{jt}$ only by a
common constant, so the cross-group premia below are unaffected.} Because $\theta^M_{jt}=\nu\,S^M_{jt}$ is read off the firm's own cost share, not fixed at an
industry-common value, it tracks the cost-share variation whose neglect biases the \citet{deloecker2012markups}
markup when factor-augmenting productivity is omitted \citep{doraszelski2019using,demirer2020production}.

\begin{figure}[htbp]
\centering
\includegraphics[width=0.72\textwidth]{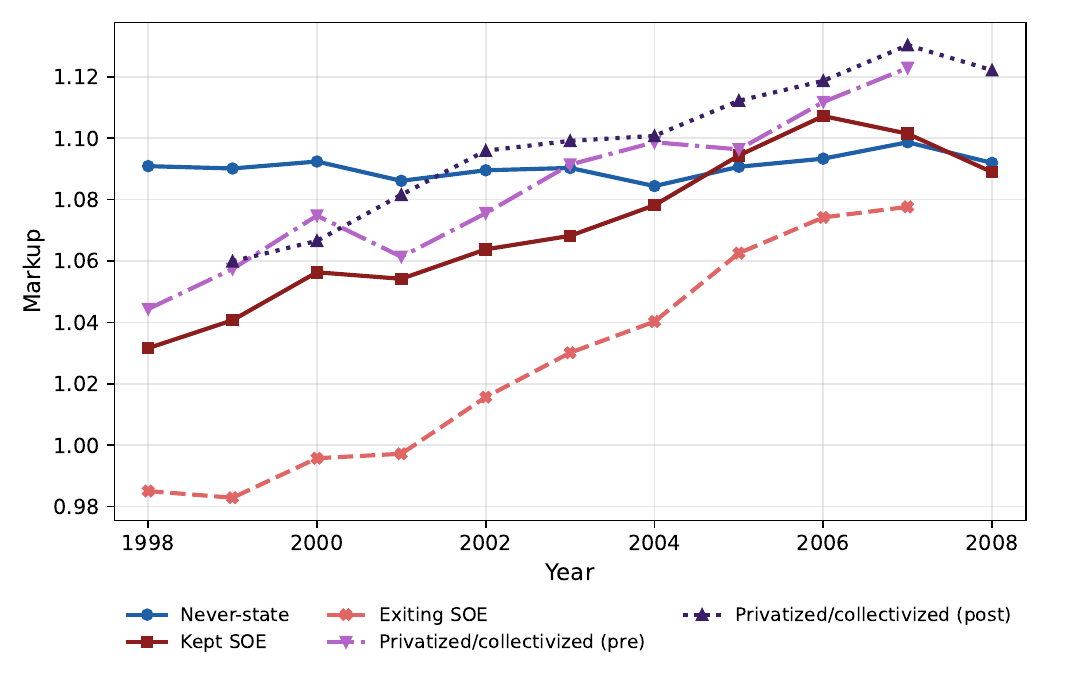}
\caption{Markup by ownership group, 1998--2008 (unweighted firm mean, markup winsorized at $1/99$). The five ownership groups, never-state (always non-SOE) firms, kept and exiting SOEs, and the privatized/collectivized firms before and after their transition, are those of Figure~\ref{fig:cost_shares}. That group pools privatization (to private/foreign) and collectivization (to collective control).}
\label{fig:markup_by_group}
\end{figure}

Across all firm-years the full-sample markup is modest (mean $1.09$, median $1.05$), in
line with the low markups reported for Chinese manufacturing over this period. Figure~\ref{fig:markup_by_group} plots $\mu$ by ownership history and Table~\ref{tab:markup_premium}
reports the corresponding premia, from a regression of the log markup on ownership-group indicators,
with kept SOEs the omitted group,
\begin{equation*}
\ln \mu_{jt} = \sum_{g}\delta^g\,D^g_{jt} + \beta_l\,l_{jt} + \beta_k\,k_{jt} + \phi_{rt} + u_{jt},
\end{equation*}
where $D^g_{jt}$ indicates firm $j$'s ownership group $g$ at $t$, $l_{jt}$ and $k_{jt}$ are log labor
and capital, $\phi_{rt}$ is a year-by-two-digit-industry ($r$) fixed effect, and $\delta^g$ is the
markup premium of group $g$ relative to the kept SOEs.\footnote{The markup uses deflated revenue, but
\citet{deloecker2012markups} show this affects only the level of the markup, not its correlation with
firm characteristics: the omitted-price bias \citep{klette1996inconsistency} loads on the constant
term, so the premia $\delta^g$ are consistent as long as the price bias is
uncorrelated with a firm's ownership given input use, which is why I include $l_{jt}$ and $k_{jt}$ as
controls.} Foreign firms carry
the largest premium, a markup $4.1\%$ above the kept SOEs. Always-private and always-collective firms sit
modestly above them ($+0.9$ and $+1.1\%$). The exiting SOEs sit far below, $-7.2\%$,
pricing near or below average cost. Privatized firms fall in between. After their transition to
private or foreign ownership they carry a markup $1.6\%$ above the kept SOEs, having stood $1.3\%$
below them in their pre-privatization years.

Markups also rise with privatization itself. The post-transition firm-years of the private or
foreign privatizers carry a markup $2.9\%$ above the same group's pre-privatization years (Panel~B).\footnote{Panel~B is a pooled contrast between post-
and pre-transition firm-years, without firm fixed effects, so privatizers observed longer after their
transition weigh more heavily in the post cell. As \citet{deloecker2012markups}
stress, this is an association rather than a causal effect.} This markup gain is specific to the dominant
private or foreign destination: privatizers reclassified as collectives show no markup change
(Panel~B, $-0.2$pp), in line with that destination being a nominal reassignment rather
than a move to market pricing. The rise is part of a broader convergence. Over
the decade the kept SOEs and the privatizers both start well below the never-state firms. The kept
SOEs close most of that gap by $2008$, while the privatizers catch up to and pass the never-state
markup (Figure~\ref{fig:markup_by_group}). The markup convergence parallels the state sector's catch-up in
factor-augmenting productivity, consistent with \citet{chen2021privatization}'s emphasis on
pricing and demand management.

\begin{table}[htbp]\centering
\begin{threeparttable}
\caption{Markup Premia by Ownership Group.}
\label{tab:markup_premium}
\begin{tabular}{lc}
\toprule
Group & $\delta^g$ (log markup) \\
\midrule
\multicolumn{2}{l}{\textit{Panel A. Ownership-group level premia}} \\
Always private & $+0.009^{***}$ \\
 & $(0.003)$ \\
Always foreign & $+0.040^{***}$ \\
 & $(0.003)$ \\
Always collective & $+0.011^{***}$ \\
 & $(0.003)$ \\
Pre-privatization SOE & $-0.013^{***}$ \\
 & $(0.004)$ \\
Post SOE$\to$Private/Foreign & $+0.016^{***}$ \\
 & $(0.004)$ \\
Post SOE$\to$Collective & $-0.014^{***}$ \\
 & $(0.005)$ \\
Exiting SOE & $-0.075^{***}$ \\
 & $(0.003)$ \\
\midrule
\multicolumn{2}{l}{\textit{Panel B. Privatization association, post $-$ pre}} \\
Post SOE$\to$Private/Foreign $-$ Pre & $+0.029^{***}$ \\
 & $(0.002)$ \\
Post SOE$\to$Collective $-$ Pre & $-0.002$ \\
 & $(0.004)$ \\
\midrule
Controls (log $L$, log $K$) & Yes \\
Year $\times$ industry FE & Yes \\
Firms & 372{,}811 \\
Observations & 1{,}659{,}599 \\
$R^2$ (within) & 0.021 \\
\bottomrule
\end{tabular}
\begin{tablenotes}\footnotesize
\item \textit{Notes:} The dependent variable is the log markup $\ln\mu_{jt}$, with $\mu_{jt}$ winsorized at the 1st/99th percentiles. Each entry is the estimated coefficient $\delta^g$ on the ownership-group indicator, i.e.\ the log-markup premium of the group over the omitted kept-SOE group; the implied markup is $100(e^{\delta^g}-1)$ percent higher (e.g.\ $+4.1\%$ for always-foreign firms). All columns include log labor, log capital, and year-by-two-digit-industry fixed effects. Standard errors clustered by firm, in parentheses; ${}^{*}\,p<0.10$, ${}^{**}\,p<0.05$, ${}^{***}\,p<0.01$. The omitted baseline contains kept-SOE firm-years only; flicker firms carry a separate unreported nuisance indicator. The sample is marginally smaller than the full analysis panel because firm-years with a missing or non-positive markup are dropped before winsorizing.
\end{tablenotes}
\end{threeparttable}
\end{table}

\section{Full Per-Industry Productivity-Process Coefficients}\label{Appendix:industry_full}
Table~\ref{tab:industry} in the main text reports the per-industry privatization effects $\beta^{w}_{d}-\pi_{d}$ by destination and timing. For completeness, Tables~\ref{tab:industry_full_K}--\ref{tab:industry_full_M} report the full set of ownership components of the productivity process $z^X_{jt}$ estimated separately within each of the ten industries: the never-SOE terms $\alpha_P,\alpha_C,\alpha_F$, the exiting-SOE term $\alpha_E$, the destination-split pre-privatization selection terms $\pi_{SP},\pi_{SC}$, and the early/late post-transition terms $\beta^{\mathrm{ea}}_{SP},\beta^{\mathrm{la}}_{SP},\beta^{\mathrm{ea}}_{SC},\beta^{\mathrm{la}}_{SC}$, for each factor-augmenting productivity, with standard errors clustered by firm. These are the per-industry version of the pooled Table~\ref{tab:gmm_priv}. The core CES parameters ($\hat\sigma,\hat\nu,\rho^X$) by industry are in Table~\ref{tab:industry_prod}.

\begin{landscape}
\begin{table}[htbp]
\centering
\begin{threeparttable}
\caption{Full Per-Industry Productivity-Process Coefficients: Capital-Augmenting Efficiency $\omega^K$.}
\label{tab:industry_full_K}
{\footnotesize\setlength{\tabcolsep}{2.5pt}\renewcommand{\arraystretch}{1.0}
\begin{tabular}{lcccccccccc}
\toprule
Industry & $\alpha_P$ & $\alpha_C$ & $\alpha_F$ & $\alpha_E$ & $\pi_{SP}$ & $\pi_{SC}$ & $\beta_{SP}^{\text{ea}}$ & $\beta_{SP}^{\text{la}}$ & $\beta_{SC}^{\text{ea}}$ & $\beta_{SC}^{\text{la}}$ \\
\midrule
Food, drink and tobacco & $\phantom{-}0.141^{***}$ & $\phantom{-}0.185^{***}$ & $-0.028$ & $-0.268^{***}$ & $-0.040$ & $\phantom{-}0.102$ & $-0.040$ & $-0.115^{***}$ & $-0.186^{***}$ & $-0.109$ \\
Textile, leather and shoes & $\phantom{-}0.074^{*}$ & $\phantom{-}0.073^{*}$ & $\phantom{-}0.033$ & $-0.141^{***}$ & $\phantom{-}0.021$ & $\phantom{-}0.132$ & $\phantom{-}0.051$ & $-0.031$ & $-0.002$ & $-0.052$ \\
Timber and furniture & $\phantom{-}0.149$ & $\phantom{-}0.173$ & $\phantom{-}0.029$ & $-0.199^{*}$ & $-0.033$ & $-0.198$ & $-0.067$ & $-0.047$ & $\phantom{-}0.183^{\dagger}$ & $\phantom{-}0.144^{\dagger}$ \\
Paper & $\phantom{-}0.343^{***}$ & $\phantom{-}0.355^{***}$ & $\phantom{-}0.239^{**}$ & $-0.182$ & $\phantom{-}0.030$ & $\phantom{-}0.081$ & $\phantom{-}0.111$ & $\phantom{-}0.043$ & $\phantom{-}0.322^{**\dagger}$ & $\phantom{-}0.129^{\dagger}$ \\
Printing & $\phantom{-}0.125^{***}$ & $\phantom{-}0.151^{***}$ & $\phantom{-}0.133^{***}$ & $-0.003$ & $\phantom{-}0.079$ & $\phantom{-}0.154$ & $\phantom{-}0.041$ & $\phantom{-}0.015$ & $-0.021^{\dagger}$ & $-0.200^{\dagger}$ \\
Chemical products & $\phantom{-}0.208^{***}$ & $\phantom{-}0.247^{***}$ & $\phantom{-}0.101^{***}$ & $-0.102^{***}$ & $\phantom{-}0.072^{**}$ & $\phantom{-}0.015$ & $\phantom{-}0.082^{**}$ & $-0.043$ & $\phantom{-}0.141^{**}$ & $\phantom{-}0.123$ \\
Non-metallic minerals & $\phantom{-}0.109^{***}$ & $\phantom{-}0.107^{***}$ & $\phantom{-}0.005$ & $-0.124^{***}$ & $\phantom{-}0.020$ & $-0.065$ & $\phantom{-}0.000$ & $-0.088^{***}$ & $-0.015$ & $-0.005$ \\
Metals and metal products & $\phantom{-}0.389^{***}$ & $\phantom{-}0.350^{***}$ & $\phantom{-}0.040$ & $-0.517^{***}$ & $-0.099$ & $\phantom{-}0.396$ & $\phantom{-}0.121$ & $\phantom{-}0.014$ & $\phantom{-}0.434^{*}$ & $\phantom{-}0.187$ \\
Machinery and transport equipment & $\phantom{-}0.168^{***}$ & $\phantom{-}0.236^{***}$ & $\phantom{-}0.070^{**}$ & $-0.319^{***}$ & $-0.033$ & $\phantom{-}0.155$ & $\phantom{-}0.029$ & $\phantom{-}0.004$ & $-0.064$ & $\phantom{-}0.123$ \\
Electronics & $\phantom{-}0.135^{***}$ & $\phantom{-}0.188^{***}$ & $\phantom{-}0.147^{***}$ & $-0.174^{***}$ & $\phantom{-}0.083$ & $-0.006$ & $\phantom{-}0.093$ & $\phantom{-}0.050$ & $\phantom{-}0.089$ & $\phantom{-}0.261^{**}$ \\
\bottomrule
\end{tabular}}
\begin{tablenotes}\footnotesize
\item \textit{Notes:} Per-industry estimates of the ownership components of $z^X_{jt}$ (Eq.~\ref{eq:z_def}), estimated separately within each industry against the kept-SOE baseline. $\pi_{SP},\pi_{SC}$ are destination-split pre-privatization terms; $\beta^{\text{ea}},\beta^{\text{la}}$ the early/late post-transition terms. Standard errors clustered by firm.
\item[$\dagger$] SOE$\to$Collective cells (early and late) in industries with under $100$ early firm-years (imprecise).
\item ${}^{*}\,p<0.10$,\ ${}^{**}\,p<0.05$,\ ${}^{***}\,p<0.01$.
\end{tablenotes}
\end{threeparttable}
\end{table}
\end{landscape}

\begin{landscape}
\begin{table}[htbp]
\centering
\begin{threeparttable}
\caption{Full Per-Industry Productivity-Process Coefficients: Labor-Augmenting Efficiency $\omega^L$.}
\label{tab:industry_full_L}
{\footnotesize\setlength{\tabcolsep}{2.5pt}\renewcommand{\arraystretch}{1.0}
\begin{tabular}{lcccccccccc}
\toprule
Industry & $\alpha_P$ & $\alpha_C$ & $\alpha_F$ & $\alpha_E$ & $\pi_{SP}$ & $\pi_{SC}$ & $\beta_{SP}^{\text{ea}}$ & $\beta_{SP}^{\text{la}}$ & $\beta_{SC}^{\text{ea}}$ & $\beta_{SC}^{\text{la}}$ \\
\midrule
Food, drink and tobacco & $\phantom{-}0.192^{***}$ & $\phantom{-}0.151^{***}$ & $\phantom{-}0.133^{***}$ & $-0.183^{***}$ & $\phantom{-}0.049$ & $\phantom{-}0.165^{**}$ & $\phantom{-}0.070^{**}$ & $\phantom{-}0.003$ & $-0.062$ & $-0.013$ \\
Textile, leather and shoes & $\phantom{-}0.068^{**}$ & $\phantom{-}0.032$ & $-0.004$ & $-0.087^{**}$ & $\phantom{-}0.030$ & $\phantom{-}0.076$ & $\phantom{-}0.041$ & $-0.004$ & $-0.069$ & $-0.101$ \\
Timber and furniture & $\phantom{-}0.067$ & $\phantom{-}0.033$ & $\phantom{-}0.009$ & $-0.214^{***}$ & $-0.008$ & $-0.045$ & $-0.042$ & $-0.009$ & $-0.039^{\dagger}$ & $\phantom{-}0.067^{\dagger}$ \\
Paper & $\phantom{-}0.219^{**}$ & $\phantom{-}0.177^{**}$ & $\phantom{-}0.264^{***}$ & $-0.281^{***}$ & $\phantom{-}0.083$ & $-0.265^{*}$ & $\phantom{-}0.051$ & $\phantom{-}0.043$ & $\phantom{-}0.059^{\dagger}$ & $\phantom{-}0.119^{\dagger}$ \\
Printing & $\phantom{-}0.150^{***}$ & $\phantom{-}0.104^{***}$ & $\phantom{-}0.130^{***}$ & $-0.224^{***}$ & $-0.027$ & $\phantom{-}0.044$ & $\phantom{-}0.011$ & $-0.007$ & $-0.025^{\dagger}$ & $\phantom{-}0.112^{\dagger}$ \\
Chemical products & $\phantom{-}0.142^{***}$ & $\phantom{-}0.102^{***}$ & $\phantom{-}0.126^{***}$ & $-0.169^{***}$ & $\phantom{-}0.003$ & $-0.008$ & $\phantom{-}0.061^{**}$ & $\phantom{-}0.008$ & $-0.006$ & $-0.029$ \\
Non-metallic minerals & $\phantom{-}0.092^{***}$ & $\phantom{-}0.040^{*}$ & $\phantom{-}0.079^{***}$ & $-0.122^{***}$ & $\phantom{-}0.030$ & $\phantom{-}0.019$ & $\phantom{-}0.021$ & $-0.034$ & $-0.008$ & $-0.076^{*}$ \\
Metals and metal products & $\phantom{-}0.066$ & $-0.061$ & $-0.095$ & $-0.484^{***}$ & $-0.117$ & $\phantom{-}0.108$ & $-0.105$ & $\phantom{-}0.009$ & $-0.054$ & $-0.506^{*}$ \\
Machinery and transport equipment & $\phantom{-}0.243^{***}$ & $\phantom{-}0.188^{***}$ & $\phantom{-}0.245^{***}$ & $-0.238^{***}$ & $\phantom{-}0.021$ & $\phantom{-}0.036$ & $\phantom{-}0.085^{***}$ & $\phantom{-}0.078^{**}$ & $-0.042$ & $-0.010$ \\
Electronics & $\phantom{-}0.181^{***}$ & $\phantom{-}0.180^{***}$ & $\phantom{-}0.209^{***}$ & $-0.158^{***}$ & $\phantom{-}0.078^{*}$ & $-0.055$ & $\phantom{-}0.106^{***}$ & $\phantom{-}0.123^{***}$ & $\phantom{-}0.084$ & $\phantom{-}0.257^{***}$ \\
\bottomrule
\end{tabular}}
\begin{tablenotes}\footnotesize
\item \textit{Notes:} Per-industry estimates of the ownership components of $z^X_{jt}$ (Eq.~\ref{eq:z_def}), estimated separately within each industry against the kept-SOE baseline. $\pi_{SP},\pi_{SC}$ are destination-split pre-privatization terms; $\beta^{\text{ea}},\beta^{\text{la}}$ the early/late post-transition terms. Standard errors clustered by firm.
\item[$\dagger$] SOE$\to$Collective cells (early and late) in industries with under $100$ early firm-years (imprecise).
\item ${}^{*}\,p<0.10$,\ ${}^{**}\,p<0.05$,\ ${}^{***}\,p<0.01$.
\end{tablenotes}
\end{threeparttable}
\end{table}
\end{landscape}

\begin{landscape}
\begin{table}[htbp]
\centering
\begin{threeparttable}
\caption{Full Per-Industry Productivity-Process Coefficients: Material-Augmenting Efficiency $\omega^M$.}
\label{tab:industry_full_M}
{\footnotesize\setlength{\tabcolsep}{2.5pt}\renewcommand{\arraystretch}{1.0}
\begin{tabular}{lcccccccccc}
\toprule
Industry & $\alpha_P$ & $\alpha_C$ & $\alpha_F$ & $\alpha_E$ & $\pi_{SP}$ & $\pi_{SC}$ & $\beta_{SP}^{\text{ea}}$ & $\beta_{SP}^{\text{la}}$ & $\beta_{SC}^{\text{ea}}$ & $\beta_{SC}^{\text{la}}$ \\
\midrule
Food, drink and tobacco & $-0.063^{***}$ & $-0.059^{***}$ & $\phantom{-}0.003$ & $\phantom{-}0.043^{***}$ & $\phantom{-}0.011$ & $-0.057^{**}$ & $-0.005$ & $\phantom{-}0.012$ & $\phantom{-}0.024$ & $-0.004$ \\
Textile, leather and shoes & $-0.052^{***}$ & $-0.047^{**}$ & $-0.011$ & $\phantom{-}0.010$ & $-0.021$ & $-0.053^{*}$ & $-0.031$ & $-0.049^{**}$ & $-0.022$ & $\phantom{-}0.003$ \\
Timber and furniture & $-0.012$ & $-0.005$ & $\phantom{-}0.012$ & $\phantom{-}0.131^{***}$ & $\phantom{-}0.072^{**}$ & $\phantom{-}0.075^{*}$ & $\phantom{-}0.075^{***}$ & $\phantom{-}0.038$ & $\phantom{-}0.040^{\dagger}$ & $-0.031^{\dagger}$ \\
Paper & $-0.160^{***}$ & $-0.154^{***}$ & $-0.107^{*}$ & $\phantom{-}0.056$ & $-0.055$ & $\phantom{-}0.001$ & $-0.035$ & $-0.107^{*}$ & $-0.113^{*\dagger}$ & $-0.150^{**\dagger}$ \\
Printing & $-0.122^{***}$ & $-0.122^{***}$ & $-0.090^{***}$ & $\phantom{-}0.107^{***}$ & $-0.013$ & $-0.116$ & $-0.005$ & $-0.009$ & $\phantom{-}0.009^{\dagger}$ & $-0.015^{\dagger}$ \\
Chemical products & $-0.081^{***}$ & $-0.099^{***}$ & $-0.027^{*}$ & $\phantom{-}0.006$ & $\phantom{-}0.002$ & $-0.055^{*}$ & $-0.003$ & $\phantom{-}0.027$ & $-0.062^{***}$ & $-0.053^{**}$ \\
Non-metallic minerals & $-0.075^{***}$ & $-0.080^{***}$ & $-0.029^{**}$ & $\phantom{-}0.030^{**}$ & $\phantom{-}0.000$ & $-0.010$ & $-0.029^{**}$ & $-0.046^{***}$ & $-0.029$ & $-0.043^{*}$ \\
Metals and metal products & $-0.167^{***}$ & $-0.148^{***}$ & $-0.050$ & $\phantom{-}0.126^{***}$ & $-0.014$ & $-0.215^{***}$ & $-0.060$ & $-0.119^{***}$ & $-0.145^{***}$ & $-0.083$ \\
Machinery and transport equipment & $-0.147^{***}$ & $-0.138^{***}$ & $-0.087^{***}$ & $\phantom{-}0.130^{***}$ & $\phantom{-}0.013$ & $-0.041$ & $-0.033^{*}$ & $-0.063^{***}$ & $-0.022$ & $-0.058$ \\
Electronics & $-0.122^{***}$ & $-0.126^{***}$ & $-0.086^{***}$ & $\phantom{-}0.057^{**}$ & $-0.049^{*}$ & $\phantom{-}0.016$ & $-0.050^{**}$ & $-0.058^{**}$ & $-0.051$ & $-0.167^{***}$ \\
\bottomrule
\end{tabular}}
\begin{tablenotes}\footnotesize
\item \textit{Notes:} Per-industry estimates of the ownership components of $z^X_{jt}$ (Eq.~\ref{eq:z_def}), estimated separately within each industry against the kept-SOE baseline. $\pi_{SP},\pi_{SC}$ are destination-split pre-privatization terms; $\beta^{\text{ea}},\beta^{\text{la}}$ the early/late post-transition terms. Standard errors clustered by firm.
\item[$\dagger$] SOE$\to$Collective cells (early and late) in industries with under $100$ early firm-years (imprecise).
\item ${}^{*}\,p<0.10$,\ ${}^{**}\,p<0.05$,\ ${}^{***}\,p<0.01$.
\end{tablenotes}
\end{threeparttable}
\end{table}
\end{landscape}

\end{document}